\documentclass[12pt]{article}

\ifx\pdfoutput\undefined
\usepackage[dvips,bookmarks]{hyperref}
\else
\usepackage{hyperref}
\fi
\hypersetup{colorlinks=false,bookmarksopen,bookmarksnumbered,citecolor=blue,
   pdfstartview=FitH}

\usepackage[dvips]{graphicx}
\usepackage{latexsym}

\usepackage{amssymb,amsfonts,amsmath}
\usepackage{graphicx} 
\usepackage{indentfirst}

 \usepackage{bbm}

\parskip=\medskipamount

\newcommand {\cA}{{\cal A}}

\newcommand {\cD}{{\cal D}}

\newcommand {\cF}{{\cal F}}

\newcommand {\cL}{{\cal L}}
\newcommand {\cM}{{\cal M}}
\newcommand {\cN}{{\cal N}}
\newcommand {\cO}{{\cal O}}

\newcommand {\cR}{{\cal R}}

\newcommand {\cT}{{\cal T}}

\newcommand {\cV}{{\cal V}}

\def\a{\alpha}
\def \bi{\bibitem}

\def\b{\beta}

\def\d{\delta}
\def\e{\epsilon}
\def\f{\phi}
\def\g{\gamma}
\def\G{\Gamma}

\def\j{\psi}

\def\m{\mu}
\def\n{\nu}
\def\o{\omega}

\def\q{\theta}

\def\s{\sigma}
\def\t{\tau}

\def\D{\Delta}
\def\F{\Phi}
\def\J{\Psi}

\def\O{\Omega}

\def\S{\Sigma}

\def\X{\Xi}
\def\tr{{\rm tr}}
\def\rd{{\rm d}}
\def\ri{{\rm i}}

\newcommand{\ad}{{\dot{\alpha}}}                           
\newcommand{\bd}{{\dot{\beta}}}                            
\newcommand{\ve}{\varepsilon}                            

\newcommand{\pa}{\partial}                           
\newcommand{\hf}{\frac12}

\newcommand{\sect}[1]{\setcounter{equation}{0}\section{#1}}

\newcommand{\be}{\begin{equation}}
\newcommand{\ee}{\end{equation}}
\newcommand{\bea}{\begin{eqnarray}}
\newcommand{\eea}{\end{eqnarray}}
\newcommand{\non}{\nonumber}
\newcommand{\1}{\underline{1}}
\newcommand{\2}{\underline{2}}

\def\dt#1{{\buildrel {\hbox{\LARGE .}} \over {#1}}}    

\def\double #1{#1{\hbox{\kern-2pt $#1$}}}

\newcommand{\hm}{{\hat{m}}}
\newcommand{\hn}{{\hat{n}}}
\newcommand{\hp}{{\hat{p}}}
\newcommand{\hq}{{\hat{q}}}
\newcommand{\ha}{{\hat{a}}}
\newcommand{\hb}{{\hat{b}}}
\newcommand{\hc}{{\hat{c}}}
\newcommand{\hd}{{\hat{d}}}
\newcommand{\he}{{\hat{e}}}

\newcommand{\hal}{{\hat{\a}}}
\newcommand{\hbe}{{\hat{\b}}}
\newcommand{\hga}{{\hat{\g}}}
\newcommand{\hde}{{\hat{\d}}}
\newcommand{\hrh}{{\hat{\rho}}}
\newcommand{\hta}{{\hat{\tau}}}

\newcommand{\halu}{{\underline{\hal}}}
\newcommand{\hbeu}{{\underline{\hbe}}}
\newcommand{\hgau}{{\underline{\hga}}}
\newcommand{\hdeu}{{\underline{\hde}}}

\newcommand{\CD}{{\nabla}}

\begin{document}

\begin{titlepage}

\begin{flushright}
December, 2007\\
\end{flushright}
\vspace{5mm}

\begin{center}
{\Large \bf  5D Supergravity and Projective Superspace}\\ 
\end{center}

\begin{center}

{\large  
Sergei M. Kuzenko\footnote{{kuzenko@cyllene.uwa.edu.au}}
and 
Gabriele Tartaglino-Mazzucchelli\footnote{gtm@cyllene.uwa.edu.au}
} \\
\vspace{5mm}

\footnotesize{
{\it School of Physics M013, The University of Western Australia\\
35 Stirling Highway, Crawley W.A. 6009, Australia}}  
~\\
\vspace{2mm}

\end{center}
\vspace{5mm}

\begin{abstract}
\baselineskip=14pt
This paper is a companion to our earlier work  \cite{KT-Msugra5D}
in which the projective superspace 
formulation for matter-coupled simple supergravity in five dimensions 
was presented. For the minimal multiplet of 5D $\cN=1$  supergravity introduced 
by Howe in 1981, we give a complete solution of the Bianchi identities. 
The geometry of curved superspace is shown to allow the existence of a large family of off-shell 
supermultiplets that can be used to describe supersymmetric matter, including vector 
multiplets and hypermultiplets. 
We formulate a manifestly locally supersymmetric 
action principle. Its natural property turns out to be the invariance under so-called projective
transformations of the auxiliary isotwistor variables.   We then demonstrate that 
the projective invariance allows one to uniquely restore the action functional in 
a Wess-Zumino  gauge. The latter action is well-suited for reducing 
the supergravity-matter systems to components.
\end{abstract}
\vspace{1cm}

\vfill
\end{titlepage}

\newpage
\renewcommand{\thefootnote}{\arabic{footnote}}
\setcounter{footnote}{0}

\tableofcontents{}
\vspace{1cm}
\bigskip\hrule


\sect{Introduction}

In our recent paper \cite{KT-Msugra5D},  the projective superspace 
formulation for matter-coupled simple supergravity in five dimensions 
was presented. 
Building on the earlier work of \cite{KT-M}, Ref.  \cite{KT-Msugra5D}
 provided the first solution to the important old problem of incorporating 
supergravity into the projective superspace approach \cite{KLR,LR1}. 
The latter  is known to be a powerful paradigm for constructing 
off-shell rigid supersymmetric theories 
with eight supercharges in $D \leq 6$ space-time dimensions, and in particular for 
the explicit construction of hyperk\"ahler metrics, see, e.g.,  \cite{HitKLR}.
In  \cite{KT-Msugra5D},  we introduced various supermultiplets to describe matter fields
coupled to supergravity, stated the locally supersymmetric action principle in the
Wess-Zumino gauge, and constructed several interesting supergravity-matter systems.

The present paper, on one hand,  is a companion to \cite{KT-Msugra5D}.
Here we derive those technical details that were stated in \cite{KT-Msugra5D}
without proof.
In particular, we show that the requirement of projective invariance allows one
to uniquely reconstruct  the locally supersymmetric action in the Wess-Zumino gauge. 
On the other hand, this paper contains an important new result. 
Specifically,  we formulate a manifestly locally supersymmetric action
that reduces to that given in \cite{KT-Msugra5D} upon imposing the 
Wess-Zumino gauge. This result completes the formal structure of 5D 
$\cN=1$ superfield supergravity.

Before turning to the technical aspects of this work, we would like to give two general 
comments.  First, five-dimensional $\cN=1$ supergravity\footnote{On historical grounds, 
5D simple ($\cN=1$) supersymmetry and supergravity are often labeled $\cN=2$.} 
\cite{Cremmer} and its matter couplings have extensively been studied
at the component level, both  in on-shell \cite{GST,GZ,CD} 
and off-shell  \cite{Zucker,Ohashi,Bergshoeff} settings. 
It is thus natural to ask: Are there still good reasons for developing superspace formulations?
We believe the answer is ``Yes.'' There are several ways to justify this claim, 
and the most practical is the following.
Unlike the component schemes developed, 
superspace approaches have the potential to offer a generating formalism 
to realize most general sigma-model couplings, and hence
to construct  general quaternionic K\"ahler manifolds.
It is instructive to discuss the situation with hypermultiplets.
In the component formulations\footnote{Refs. \cite{Bergshoeff} deal with on-shell
hypermultiplets only.}
of  \cite{Zucker,Ohashi}, one makes use of an off-shell realization 
for the hypermultiplet with finitely many auxiliary fields and an intrinsic central charge. 
As is well-known, it is the presence of  central charge which makes it impossible
to cast general quaternionic K\"ahler couplings in terms of such 
off-shell hypermultiplets. 
On the other hand, the projective superspace approach offers
nice off-shell formulations without central charge. 
Specifically, there are infinitely many  off-shell realizations with {\it finitely many auxiliary fields}
 for a neutral hypermultiplet (they are the called $O(2n)$ multiplets,  where $n=2, 3\dots$, 
 following the terminology 
of \cite{G-RLRvUW}), and a unique formulation for a charged hypermultiplet 
with {\it infinitely many auxiliary fields}
(the so-called polar hypermultiplet).

Our second comment concerns the  choice made in this paper 
to use the projective superspace setting 
to formulate supergravity-matter systems. 
Why not harmonic superspace \cite{GIKOS,GIOS}?
As is known, both  approaches can be used to describe supersymmetric theories 
with eight supercharges in $D \leq 6$ space-time dimensions. 
There are, however,  two major differences between them: 
(i) the structure of off-shell supermultiplets used; 
and (ii) the supersymmetric action principle chosen. 
It is due to these differences that the two approaches are complementary 
to each other in some respects. From the point of view of supergravity theories 
with eight supercharges in $D \leq 6$ space-time dimensions,  
harmonic superspace offers powerful prepotential formulations
\cite{SUGRA-har,Sokatchev}.
On the other hand, as  will be shown in this paper, 
projective superspace is ideal  for developing covariant geometric formulations 
for supergravity-matter systems, 
 similar to the famous Wess-Zumino approach for 
 4D $\cN=1$ supergravity \cite{WZ-s}. 
The point is  that projective superspace is a robust scheme for supersymmetric model-buliding, 
see, e.g., \cite{AKL} for the recent construction of hyperk\"ahler 
metrics on cotangent bundles of Hermitian symmetric spaces. 

This paper is organized as follows. In section 2 we provide a complete solution 
of the Bianchi identities for the superspace geometry corresponding to 
the minimal 5D $\cN=1$ supergravity multiplet \cite{Howe5Dsugra}. In section 3
we formulate, following  \cite{KT-Msugra5D}, off-shell projective supermultiplets, 
and then construct a manifestly locally supersymmetric action.
Section 4 is devoted to the technicalities of the Wess-Zumino gauge for supergravity.
Section 5 demonstrates that the locally supersymmetric action in the Wess-Zumino gauge
is uniquely determined from the requirement of projective invariance. 
Our 5D conventions and useful identities are collected in the appendix. 

\sect{Superspace geometry of the minimal supergravity multiplet}
\label{SectionSugraGeometry}

In this section we present a complete solution to the Bianchi identities for
the constraints on the superspace torsions that were introduced by Howe\footnote{The
choice of the constraints given in   \cite{Howe5Dsugra} was motivated by the structure of the
5D $\cN=1$ supercurrent   \cite{HL}.}
in 1981 \cite{Howe5Dsugra} and 
correspond to the so-called minimal 5D $\cN=1$ supergravity multiplet.\footnote{This 
supermultiplet was re-discovered almost twenty years later by Zucker \cite{Zucker}
who essentially elaborated the component implications of the superspace formulation 
given in  \cite{Howe5Dsugra}.}
The results of this section were used in 
 \cite{KT-Msugra5D} without proof.

Let $z^{\hat{M}}=(x^{\hm},\q^{\hat{\mu}}_i)$
be local bosonic ($x$) and fermionic ($\q$) 
coordinates parametrizing  a curved five-dimensional $\cN=1$  superspace
$\cM^{5|8}$,
where $\hm=0,1,\cdots,4$, $\hat{\mu}=1,\cdots,4$, and  $i=\1,\2$.
The Grassmann variables $\q^{\hat{\mu}}_i$
are assumed to obey the standard pseudo-Majorana reality condition
$(\q^{\hat{\mu}}_i)^* = \q_{\hat{\mu}}^i =\ve_{\hat{\m} \hat{\n}}\,  \ve^{ij} \, \q^{\hat{\nu}}_j  $
(see the appendix for our 5D  notation and conventions).
${}$Following \cite{Howe5Dsugra}, the tangent-space group
is chosen to be  ${\rm SO}(4,1)\times {\rm SU}(2)$,
and the superspace  covariant derivatives 
$\cD_{\hat{A}} =(\cD_{\hat{a}}, \cD_{\hat{\a}}^i) \equiv (\cD_{\hat{a}}, \cD_{\halu})$
have the form 
\bea
\cD_{\hat{A}}&=&
E_{\hat{A}} + \O_{\hat{A}} + \F_{\hat{A}}
+V_{\hat{A}}Z~.
\label{CovDev}
\eea
Here $E_{\hat{A}}= E_{\hat{A}}{}^{\hat{M}}(z) \,\pa_{\hat{M}}$ is the supervielbein, 
with $\pa_{\hat{M}}= \pa/ \pa z^{\hat{M}}$,
\bea
\O_{\hat{A} }= \hf \,\O_{\hat{A}}{}^{\hb\hc}\,M_{\hb\hc}
= \O_{\hat{A}}{}^{\hbe\hga}\,M_{\hbe\hga}~,\qquad 
M_{\ha\hb}=-M_{\hb\ha}~, \quad M_{\hal\hbe}=M_{\hbe\hal}
\eea
is the Lorentz connection, 
\bea
\F_{\hat{A}} = \F^{~\,kl}_{\hat{A}}\,J_{kl}~, \qquad
J_{kl}=J_{lk}
\eea
is the SU(2)-connection, and $Z$ the central-charge generator, $[Z, \cD_{\hat{A}}]=0$.
The Lorentz generators with vector indices ($M_{\ha\hb}$) and spinor indices
($M_{\hal\hbe}$) are related to each other by the rule:
$M_{\ha\hb}=(\S_{\ha\hb})^{\hal\hbe}M_{\hal\hbe}$ 
(for more details, see the appendix).
The generators of ${\rm SO}(4,1)\times {\rm SU}(2)$
act on the covariant derivatives as follows:\footnote{The operation of
(anti)symmetrization of $n$ indices 
is defined to involve a factor $(n!)^{-1}$.}
\bea
{[}J^{kl},\cD_{\hal}^i{]}
= \ve^{i(k} \cD^{l)}_{\hat \a}~,~~~
{[}M_{\hal\hbe},\cD_{\hga}^k{]}
=\ve_{\hga(\hal}\cD^k_{\hbe)}~,~~~
{[}M_{\ha\hb},\cD_{\hc}{]}
=2\eta_{\hc[\ha}\cD_{\hb]}~,
\label{generators}
\eea
where $J^{kl} =\ve^{ki}\ve^{lj} J_{ij}$.

The supergravity gauge group is generated by local transformations
of the form 
\be
\cD_{\hat{A}} \to \cD'_{\hat{A}}  ={\rm e}^{K  }\, \cD_{\hat{A}}\, {\rm e}^{-K  } ~,
\qquad K = K^{\hat{C}}(z) \cD_{\hat{C}} +\hf K^{\hat c \hat d}(z) M_{\hat c \hat d}
+K^{kl}(z) J_{kl}  +\t(z) Z~,
\label{tau}
\ee
with all the gauge parameters being neutral with respect to the central charge $Z$,
obeying natural reality conditions, and otherwise  arbitrary. 
Given a tensor superfield $U(z)$, with its indices suppressed, 
it transforms as follows:
\bea
U \to U' = {\rm e}^{K  }\, U~.
\eea

The covariant derivatives obey (anti)commutation relations of the general form 
\bea
{[}\cD_{\hat{A}},\cD_{\hat{B}}\}&=&T_{\hat{A}\hat{B}}{}^{\hat{C}}\cD_{\hat{C}}
+\hf R_{\hat{A}\hat{B}}{}^{\hat{c}\hat{d}}M_{\hat{c}\hat{d}}
+R_{\hat{A}\hat{B}}{}^{kl}J_{kl}
+F_{\hat{A}\hat{B}}Z~,
\label{algebra}
\eea
where $T_{\hat{A}\hat{B}}{}^{\hat{C}}$ is the torsion, 
$R_{\hat{A}\hat{B}}{}^{kl}$ and $R_{\hat{A}\hat{B}}{}^{\hat{c}\hat{d}}$  
the SU(2)- and SO(4,1)-curvature tensors, respectively, 
and $F_{\hat{A}\hat{B}}$  the central charge field strength.

The Bianchi identities are: 
\bea
\sum_{(\hat{A}\hat{B}\hat{C})}
{[}\cD_{\hat{A}},{[}\cD_{\hat{B}},\cD_{\hat{C}}\}\}
~=~0~,~~~
\label{Bianchi0}
\eea
with the graded cyclic sum assumed.
The Bianchi identities are equivalent to  the following equations 
on the torsion and curvature tensors:
\begin{subequations}
\bea
0=\sum_{(\hat{A}\hat{B}\hat{C})}
\Big(R_{\hat{A}\hat{B}}{}_{\hat{C}}{}^{\hat{D}}
-\cD_{\hat{A}}T_{\hat{B}\hat{C}}{}^{\hat{D}}
+T_{\hat{A}\hat{B}}{}^{\hat{E}}T_{\hat{E}\hat{C}}{}^{\hat{D}} \Big)~,
~~~~~~~~~~~~~~~~~~~~~~~~~
\label{Bianchi01}
\\
0=
\sum_{(\hat{A}\hat{B}\hat{C})}
\Big(\cD_{\hat{A}}R_{\hat{B}\hat{C}}{}^{kl}
-T_{\hat{A}\hat{B}}{}^{\hat{D}}R_{\hat{D}\hat{C}}{}^{kl}\Big)
~,~~~
0=
\sum_{(\hat{A}\hat{B}\hat{C})}
\Big(\cD_{\hat{A}}R_{\hat{B}\hat{C}}{}^{\hrh\hat{\tau}}
-T_{\hat{A}\hat{B}}{}^{\hat{D}}R_{\hat{D}\hat{C}}{}^{\hrh\hat{\tau}}\Big)~,~~~
\label{Bianchi03}
\\
0=
\sum_{(\hat{A}\hat{B}\hat{C})}
\Big(\cD_{\hat{A}}F_{\hat{B}\hat{C}}-T_{\hat{A}\hat{B}}{}^{\hat{D}}F_{\hat{D}\hat{C}}\Big)~,
~~~~~~~~~~~~~~~~~~~~~~~~~~~~~~~~~~
\label{Bianchi04}
\eea
\end{subequations}
where\footnote{The reader  should keep in mind 
that we often use the condensed notation: 
$A_{\halu}\equiv A_\hal^i$ and $A^\halu\equiv A^\hal_i$.}
\begin{subequations}
\bea
&R_{\hat{A}\hat{B}}{}_{\hat{C}}{}^{\hat{D}}\equiv
R_{\hat{A}\hat{B}}{}^{\hrh\hat{\tau}}(M_{\hrh\hat{\tau}})_{\hat{C}}{}^{\hat{D}}
+R_{\hat{A}\hat{B}}{}^{kl}(J_{kl})_{\hat{C}}{}^{\hat{D}}
~,\\
&
{[}M_{\hde\hrh},\cD_{\hat{A}}{]}\equiv(M_{\hde\hrh})_{\hat{A}}{}^{\hat{B}}\cD_{\hat{B}} ~,~~~
{[}J_{kl},\cD_{\hat{A}}{]}\equiv (J_{kl})_{\hat{A}}{}^{\hat{B}}\cD_{\hat{B}}
~,\\
&
(M_{\hrh\hta})_{\halu}{}^{\hbeu}=\d^i_j \ve_{\hal(\hrh}\d_{\hta)}^\hbe
~,~~
(M_{\hrh\hta})_{\hat{a}}{}^{\hat{b}}=(\S_\ha{}^{\hb})_{\hrh\hta}~,~~~
(J_{kl})_{\halu}{}^{\hbeu}=-\d_\hal^\hbe\d^i_{(k}\ve_{l)j}~,
\eea
\end{subequations}
with the other components of 
$(M_{\hrh\hta})_{\hat{C}}{}^{\hat{D}}$ and $(J_{kl})_{\hat{C}}{}^{\hat{D}}$
being equal to zero.

Similar to the well-known case of four-dimensional $\cN=1$ 
supergravity (see \cite{WB,GGRS,BK} for comprehensive reviews), 
the geometric superfields in (\ref{CovDev}) contain too many 
component fields to describe an irreducible supergravity multiplet. 
This can be cured by imposing covariant algebraic constraints on 
the geometry of superspace.  In accordance with a  theorem due to Dragon
\cite{Dragon}, it is sufficient to impose constraints on the torsion, 
since the curvature is  completely determined in terms 
of the torsion in supergravity theories formulated in superspace.

As demonstrated  in  \cite{Howe5Dsugra},  in order to realize the minimal supergravity 
multiplet in the above framework, one has to impose 
the following constraints on 
various components of the torsion
of dimensions 0, 1/2 and 1:
\begin{subequations}
\bea
T_{\hat \a}^i{}_{\hat \b}^j{}^{\hc}=-2\ri\, \ve^{ij}(\Gamma^{\hc})_{\hal\hbe}~, \qquad
F_{\hat \a}^i{}_{\hat  \b}^j=-2\ri\,
\ve^{ij}\ve_{\hal\hbe}~,  \qquad && \mbox{(dimension 0)} 
\label{constr-0} \\
T_{\hat \a}^i{}_{\hat \b}^j{}{}^{\hat \g}_k=
T_{\hat \a}^i{}_{\hb}{}^{\hc}=
F_{\hat \a}^i{}_{\hb}=0~,  \quad \qquad \qquad ~~~~&& \mbox{(dimension 1/2)} 
\label{constr-1/2} \\
T_{\ha\hb}{}^{\hc}~=~T_{\ha}{}_{\hbe}^l{\,}^{\hbe}_{(j}\,
\ve_{k)l}~=~0~. \quad \qquad 
\qquad~~~~&& \mbox{(dimension 1)}
\label{constr-1}
\eea
\end{subequations}
Under these constraints, the Bianchi identities (\ref{Bianchi01}--\ref{Bianchi04})
become non-trivial equations that have to be solved in order to determine 
the non-vanishing components of the torsion.

\subsection{The algebra of covariant derivatives}

In this subsection, we summarize the results of the solution to  
the Bianchi identities based on the constraints (\ref{constr-0})--(\ref{constr-1}), 
while the technical details will be given 
in the remainder of this section.

The algebra of covariant derivatives has the form \cite{KT-Msugra5D}
\begin{subequations}
\bea
\big\{ \cD_{\hal}^i , \cD_{\hbe}^j \big\} &=&-2 \ri \,\ve^{ij}\cD_{\hal\hbe}
-2\ri \, \ve^{ij}\ve_{\hal\hbe}Z
\non\\
&&
+3\ri \, \ve_{\hal\hbe}\ve^{ij}S^{kl}J_{kl}
-2\ri(\S^{\ha\hb})_{\hal\hbe}
\Big(F_{\ha\hb}+N_{\ha\hb}\Big)J^{ij}
\non\\
&&
-\ri \,\ve_{\hal\hbe}\ve^{ij}F^{\hc\hd}M_{\hc\hd}
+{\ri\over 4} \ve^{ij}\ve^{\ha\hb\hc\hd\he}N_{\ha\hb}(\G_\hc)_{\hal\hbe}M_{\hd\he}
+4\ri \,S^{ij}M_{\hal\hbe}
~,
\label{covDev2spinor-} 
\eea
\bea
{[}\cD_\ha,\cD_{\hbe}^j{]}&=&
{1\over 2}(\Gamma_{\hat{a}})_{\hbe}{}^{\hga}S^j{}_k\cD_{\hga}^k
-{1\over 2}\,F_{\ha\hb}(\Gamma^{\hat{b}})_{\hbe}{}^{\hga}\cD_{\hga}^j
-{1\over 8}\,\ve_{\ha\hb\hc\hd\he}N^{\hd\he}(\Sigma^{\hb\hc})_{\hbe}{}^{\hga}\cD_{\hga}^j
\non\\
&&
+\Big(-3\ve^{jk}\X_{\ha\hbe}{}^{l}
+{5\over 4} (\G_{\ha})_{\hbe}{}^{\hal}\ve^{jk}\cF_{\hal}{}^{l}
-{1\over 4}(\G_\ha)_{\hbe}{}^{\hal}\ve^{jk}\cN_\hal{}^{l}\Big) J_{kl}
\non\\
&&
+\Big(\,
\hf (\G_\ha)_{\hbe}{}^\hrh\cD_\hrh^{j}F^{\hc\hd}
-\hf (\G^\hc)_{\hbe}{}^\hrh\cD_\hrh^{j}F^{\hd}{}_{\ha}
+\hf (\G^\hd)_{\hbe}{}^\hrh\cD_\hrh^{j}F^{\hc}{}_{\ha}
\Big)
M_{\hc\hd}~,
\label{covDev2spinor-2}
\eea
\bea
{[}\cD_\ha,\cD_\hb{]}&=&
{\ri\over 2} \Big(\cD^\hga_kF_{\ha\hb}\Big)\cD_\hga^k 
-{\ri\over 8}\Big(\cD^{\hga (k} \cD_\hga^{l)}F_{\ha\hb}\Big)J_{kl}
+F_{\ha\hb}Z
\non\\
&&
+\Big(\,{1\over 4}\ve^{\hc\hd}{}_{\hm\hn{[}\ha}\cD_{\hb{]}}N^{\hm\hn}
+\hf\d^{\hc}_{{[}\ha}N_{\hb{]}\hm}N^{\hd\hm}
-{1\over 4}N_\ha{}^\hc N_\hb{}^{\hd}
-{1\over 8}\d_\ha^{\hc}\d^{\hd}_{\hb}N^{\hm\hn}N_{\hm\hn}
\non\\
&&~~~
+{\ri\over 8}(\S^{\hc\hd})^{\hga\hde}\cD_{\hga}^k \cD_{\hde k}F_{\ha\hb}
-F_{\ha}{}^{\hc}F_{\hb}{}^{\hd}
+{1\over 2}\d_\ha^{\hc}\d^{\hd}_{\hb}S^{ij}S_{ij}\Big)
M_{\hc\hd}~.
\label{covDev2spinor-3}
\eea
\end{subequations}

The components of the torsion in (\ref{covDev2spinor-})--(\ref{covDev2spinor-3})
obey further constraints implied by the Bianchi identities,  
some of which 
can be conveniently  expressed in terms of 
the three irreducible components of  $\cD_\hga^kF_{\hal\hbe}$: a completely 
symmetric third-rank tensor $W_{\hal\hbe\hga}{}^k$,
a gamma-traceless spin-vector $\X_{ \ha \,\hga}{}^k$ 
and a spinor $\cF_{\hga}{}^k$.
These components originate as follows:
\bea
&\cD_\hga^kF_{\hal\hbe}=
W_{\hal\hbe\hga}{}^k
+\X_{\hga(\hal\hbe)}{}^k
+\ve_{\hga(\hal}\cF_{\hbe)}{}^k
~,~~~\non\\
&\X_{\hga\hal\hbe}{}^k=(\G_\ha)_{\hga\hal}\X^\ha{}_{\hbe}{}^k~,~~~
(\G^\ha)_\hal{}^\hbe\X_{\ha\hbe}{}^i=0~,~~~
W_{\hal\hbe\hga}{}^k=W_{( \hal\hbe\hga )}{}^k~.
\label{decoDF1}
\eea
It is useful to have eq. (\ref{decoDF1}) rewritten in the equivalent form
($W_{\ha\hb\hga}{}^k=(\S^{\ha\hb})^{\hal\hbe}W_{\hal\hbe\hga}{}^k$)
\bea
&\cD_\hga^kF_{\ha\hb}\,=\,
W_{\ha\hb\hga}{}^k
+2(\G_{{[}\ha})_{\hga}{}^{\hde}\X_{\hb{]}\hde}{}^k
+(\S_{\ha\hb})_\hga{}^\hde \cF_{\hde}{}^k~,~~~
(\G^\ha)_\hal{}^\hga W_{\ha\hb\hga}{}^i=0~.
\label{decoDF2}
\eea
The dimension 3/2  Bianchi identities are as folllows:
\begin{subequations}
\bea
\cD_\hga^kN_{\hal\hbe}&=&
-W_{\hal\hbe\hga}{}^k
+2\,\X_{\hga(\hal\hbe)}{}^k
+\ve_{\hga(\hal}\cN_{\hbe)}{}^k~,
\label{N-or-Bianchi}\\
\cD_{\hbe}^{k}S^{jl}
&=&
{1\over 10}(\S_{\ha\hb})_\hbe{}^\hde
\ve^{k(j}\cD_{\hde}^{l)}
\Big(3F^{\ha\hb}+N^{\ha\hb}\Big)
\,=\,
-\hf \ve^{k(j}\Big(3\cF_\hbe{}^{l)}+\cN_{\hbe}{}^{l)}\Big)~.
\label{S-Bianchi} 
\eea
\end{subequations}
Equation (\ref{N-or-Bianchi}) can be equivalently expressed in the form
\bea
\cD_\hga^kN_{\ha\hb}&=&
-W_{\ha\hb\hga}{}^k
+4(\G_{{[}\ha})_{\hga}{}^{\hde}\X_{\hb{]}\hde}{}^k
+(\S_{\ha\hb})_\hga{}^\hde \cN_{\hde}{}^k~.
\label{N-or-Bianchi2}
\eea

The  dimension 2 Bianchi identities  are:
\begin{subequations}
\bea
\cD^{(k}_{{[}\hbe}\cN_{\hde{]}}{}^{l)}
&=&
-\cD^{(k}_{{[}\hbe}\cF_{\hde{]}}{}^{l)}
-{3\over 4}\cD^{\hga(k} \X_{\hbe\hde\hga}{}^{l)}
~,~~~~~~
\label{dim2-1}\\
(\G_{\ha})^{\hal\hbe}\cD^k_\hal\cN_{\hbe k}
&=&
(\G_{\ha})^{\hal\hbe}\cD_{\hal}^k \cF_{\hbe k}
+4\cD^{\hal}_{k}\X_{\ha\hal}{}^{k}
-{4\ri\over 3}\ve_{\ha\hm\hn\hp\hq}\big(N^{\hm\hn} N^{\hp\hq}+F^{\hm\hn}F^{\hp\hq}\big)
~,~~~~~~~~~~
\label{dim2-2}\\
(\S^{\ha\hb})^{\hal\hbe}\cD^k_\hal\cN_{\hbe k}&=&
-5(\S^{\ha\hb})^{\hal\hbe}\cD_{\hal}^k \cF_{\hbe k}
+6(\G^{[\ha})^{\hal\hbe}\cD_{\hal}^k \X^{\hb]}{}_{\hbe k}
+16\ri F_{\hm}{}^{[\ha}N^{\hb]\hm}~,
\label{dim2-3}\\
\cD_{{[}\hal}^{(k} W_{\hbe{]}\hga\hde}{}^{l)}&=&
{1\over 2}\ve_{\hal\hbe}\cD^{(k}_{(\hga}\cF_{\hde)}{}^{l)}
-{1\over 2}\cD^{(k}_{{[}\hal}\ve_{\hbe{]}(\hga}\cF_{\hde)}{}^{l)}
-{1\over 2}\cD^{(k}_{(\hga}\ve_{\hde){[}\hal}\cF_{\hbe{]}}{}^{l)}
\non\\
&&
+{3\over 4}\ve_{\hal\hbe}\cD^{\hrh(k}\X_{\hrh(\hga\hde)}{}^{l)}
-{3\over 16}\ve_{\hga{[}\hal}\cD^{\hrh(k}\X_{\hbe{]}\hde\hrh}{}^{l)}
-{3\over 16}\ve_{\hde{[}\hal}\cD^{\hrh(k}\X_{\hbe{]}\hga\hrh}{}^{l)}
\non\\
&&
-\cD_{{[}\hal}^{(k}\X_{\hbe{]}(\hga\hde)}{}^{l)}
+2\ri\Big(\ve_{\hal\hbe}N_{\hga\hde}
-\ve_{\hga{[}\hal}N_{\hbe{]}\hde}
-\ve_{\hde{[}\hal}N_{\hbe{]}\hga}
\Big)S^{kl}~,
\label{dim2-4}\\
0&=&\cD_\ha F_{\hb\hc}
+\cD_\hb F_{\hc\ha}
+\cD_\hc F_{\ha\hb}~.
\label{2BianchiF}
\eea
\end{subequations}
Note that eq. (\ref{2BianchiF}) can equivalently be rewritten as
\bea
0&=&
\cD_k^\hal W^{\ha\hb}{}_{\hal}{}^{k}
+2\ve^{\ha\hb\hc\hd\he}(\S_{\hc\hd})^{\hal\hbe}\cD^k_\hal\X_{\he\hbe k}
-3(\S^{\ha\hb})^{\hal\hbe}\cD^k_\hal\cF_{\hbe k}
+16\ri F_{\hc}{}^{[\ha}N^{\hb]\hc}~.
\label{dim2-5}
\eea

\subsection{Solving the Bianchi identities: dimension 1}

Now, we turn to solving 
the Bianchi identities (\ref{Bianchi01})--(\ref{Bianchi04})
based on the constraints (\ref{constr-0})--(\ref{constr-1}).
It is standard and useful to organize the analysis 
in accordance with  the increasing  dimension 
of the identities involved (from dimension 1/2 to 3).

The important simplification is that 
it is sufficient to analyze only the Bianchi identities (\ref{Bianchi01}) and 
(\ref{Bianchi04}), due to Dragon's second theorem \cite{Dragon}. 
The latter states that all the equations  (\ref{Bianchi03}) are identically satisfied, 
provided (\ref{Bianchi01}) and  
(\ref{Bianchi04}) hold.

${}$For dimension $1/2$, the relations (\ref{Bianchi01}) 
with ($\hat{A}=\halu,~\hat{B}=\hbeu,~\hat{C}=\hgau,~\hat{D}=\hd$) and  
(\ref{Bianchi04}) with ($\hat{A}=\halu,~\hat{B}=\hbeu,~\hat{C}=\hgau$) 
are identically satisfied.

${}$For dimension 1, 
there occur several Bianchi identities that originate from 
eqs. (\ref{Bianchi01}) and (\ref{Bianchi04}).
Setting ($\hat{A}=\ha,~\hat{B}=\hbeu,~\hat{C}=\hgau,~\hat{D}=\hd$)
in (\ref{Bianchi01})  gives
\bea
0&=&R_{\hbe}^j{}_{\hga}^k{}_\ha{}^\hd
+2\ri T_{\ha}{}_{\hbe}^j{}^{\hrh k}(\G^\hd)_{\hrh\hga}
+2\ri T_\ha{}_{\hga}^k{}^{\hrh j}(\G^\hd)_{\hrh\hbe}~,
\label{Bianchi-1-1}
\eea
while the choice
($\hat{A}=\halu,~\hat{B}
=\hbeu,~\hat{C}=\hgau,~\hat{D}=\hdeu$)
leads to
\bea
0&=&\,
R_{\hal}^i{}_{\hbe}^j{}_\hga{}^{\hde}\d_{l}^{k}
+R_{\hbe}^j{}_{\hga}^k{}_\hal{}^{\hde}\d_{l}^{i}
+R_{\hga}^k{}_{\hal}^i{}_\hbe{}^{\hde}\d_{l}^{j}
+R_{\hal}^i{}_{\hbe}^j{}{}^{k}{}_l\d_{\hga}^{\hat{\d}}
+R_{\hbe}^j{}_{\hga}^k{}{}^{i}{}_l\d_{\hal}^{\hat{\delta}}
+R_{\hga}^k{}_{\hal}^i{}{}^{j}{}_l\d_{\hbe}^{\hat{\d}}
\non\\
&&
-\,2\ri \ve^{ij}(\Gamma^{\hat{e}})_{\hal\hbe}T_{\he}{}_{\hga}^k{}_{}{}^\hde_l
-2\ri \ve^{jk}(\Gamma^{\hat{e}})_{\hbe\hga}T_{\he}{}_{\hal}^i{}_{}{}^\hde_l
-2\ri \ve^{ki}(\Gamma^{\hat{e}})_{\hga\hal}T_{\he}{}_{\hbe}^j{}_{}{}^\hde_l
~.
\label{Bianchi-1-2}
\eea
Choosing  ($\hat{A}=\ha,~\hat{B}=\hbeu,~\hat{C}=\hgau$) 
in  (\ref{Bianchi04})  gives
\bea
0&=&T_{\ha}{}_{\hbe}^j{}_{\hga}^{ k}
+\ve^{jk}(\G^{\hd})_{\hbe\hga}F_{\hd\ha}
+T_\ha{}_{\hga}^k{}_{\hbe}^j~.
\label{BianchiF12}
\eea

Eq. (\ref{BianchiF12}) 
implies that the dimension $1$ torsion 
can be represented in the form:
\bea
&&T_{\ha}{}_{\hbe}^j{}_{\hga}^{k}~=~
{1\over 2}\,\ve^{jk}(\Gamma^{\hat{b}})_{\hbe\hga}F_{\ha\hb}
-{1\over 4}\ve^{jk}(\Sigma^{\hb\hc})_{\hbe\hga}T_{1\ha\hb\hc}
+{1\over 2}(\Gamma^{\hat{b}})_{\hbe\hga}T_{1\ha\hb}{}^{jk}
-{1\over 4}\ve_{\hbe\hga}T_{1\ha}{}^{jk}~,
\label{T1-1}
\eea
where 
\be
T_{1\ha}{}^{jk}=T_{1\ha}{}^{kj}~,\qquad
T_{1\ha\hb}{}^{jk}=T_{1\ha\hb}{}^{kj}~,
\qquad T_{1\ha\hb\hc}=-T_{1\ha\hc\hb}~.
\ee

Equation (\ref{Bianchi-1-1}) expresses the dimension $1$ Lorentz curvature 
in terms of the torsion
\bea
R_{\hal}^i{}_{\hbe}^j{}^{\hc\hd}
&=&
-2\ri T^{\hc}{}{}_{\hal}^i{}^{\hrh j}(\G^\hd)_{\hrh\hbe}
-2\ri T^\hc{}_{\hbe}^j{}^{\hrh i}(\G^\hd)_{\hrh\hal}~.
\label{R-1-0-0-1}
\eea
Since 
$R_{\hal}^i{}_{\hbe}^j{}^{\hc\hd}=-R_{\hal}^i{}_{\hbe}^j{}^{\hd\hc}$, 
the following equation occurs
\bea
0&=&(\G^{(\ha})_{\hrh\hga}T^{\hd)}{}_{\hbe}^j{}^{\hrh k}
+(\G^{(\ha})_{\hrh\hbe}T^{\hd)}{}_{\hga}^k{}^{\hrh j}
={1\over 2}(\G^{\hc})_{\hbe\hga}\ve^{jk}T_{1}{}^{(\ha\hd)}{}_{\hc}
-2(\S^{\hat{b}(\ha})_{\hbe\hga}T_{1}{}^{\hd)}{}_{\hb}{}^{jk}
~.~~~
\eea
This holds if and only if 
$T_{1\ha\hb}{}^{kl}$ and $T_{1\ha\hb\hc}$ have the form:
\bea
T_{1\ha\hb}{}^{ij}
&=&
{1\over 5}\eta_{\ha\hb}\eta^{\hm\hn}T_{1\hm\hn}{}^{ij}\equiv \eta_{\ha\hb}S{}^{ij}~,\qquad
S{}^{ij} = S{}^{ji}~, \non \\
T_{1\ha\hb\hc}&=&-T_{1\hb\ha\hc}\equiv N_{\ha\hb\hc}~, 
\qquad N_{\ha\hb\hc}=N_{[\ha\hb\hc]}~,
\eea
for some symmetric tensor $S^{ij}$ obeying the reality condition 
$\overline{S^{ij}}= S_{ij}$,
and a  completely antisymmetric
real tensor $N_{\ha\hb\hc}$.
As a result, the Lorentz curvature (\ref{R-1-0-0-1}) takes  the form:
\bea
R_{\hal}^i{}_{\hbe}^j{}_{}{}^{\hc\hd}&=&
-\ri\ve_{\hal\hbe}\ve^{ij}F^{\hc\hd}
-{\ri\over 2} \ve^{ij}N{}^{\hc\hd\he}(\G_\he)_{\hal\hbe}
+{4\ri}S^{ij}(\S^{\hc\hd})_{\hal\hbe}~.
\label{R-1-Lorentz}
\eea

Let us now turn to eq. (\ref{Bianchi-1-2}). 
Taking  the trace over the indices $\hga$ and $\hde$, 
one derive the following equation for the 
SU(2)-curvature:
\bea
4R_{\hal}^i{}_{\hbe}^j{}{}^{kl}
+R_{\hal}^k{}_{\hbe}^j{}{}^{il}
+R_{\hal}^i{}_{\hbe}^k{}{}^{jl}&=&
\D_\hal^i{}_{\hbe}^j{}^{kl}
\label{R-1-0}
~,
\eea
with 
\be
\D_\hal^i{}_{\hbe}^j{}^{kl}=
15\ri\ve_{\hal\hbe}\ve^{ij}S^{kl}
+\frac{5\ri}{2}(\G^\hc)_{\hal\hbe}\ve^{ij}T_{1\hc}{}^{kl}
+\ri(\S^{\hd\he})_{\hal\hbe}\ve^{k(i}\ve^{j)l}\Big(6F_{\hd\he}+\ve_{\hd\he\ha\hb\hc}N^{\ha\hb\hc}\Big)
~.
\ee
Here we have used 
the explicit expressions for  the dimension 1 torsion and for the Lorentz curvature 
in terms of $S^{kl}$, $T_{1\ha}{}^{jk}$ and $N_{\ha\hb\hc}$. 

Equation (\ref{R-1-0}) allows us to express 
$R_{\hal}^i{}_{\hbe}^j{}{}^{kl}$ in terms of 
$\D_\hal^i{}_{\hbe}^j{}^{kl}$, and the result is
\be
R_\hal^i{}_{\hbe}^j{}_{}{}^{kl}=
{1\over 90}\Big(
26\D_\hal^i{}_{\hbe}^j{}^{kl}
-\D_\hal^j{}_{\hbe}^i{}^{kl}
+2\D_\hal^k{}_{\hbe}^i{}^{jl}
-7\D_\hal^k{}_{\hbe}^j{}^{il}
-7\D_\hal^i{}_{\hbe}^k{}^{jl}
+2\D_\hal^j{}_{\hbe}^k{}^{il}
\Big)~.
\label{R-1-su2-0}
\ee
It is useful to introduce the Hodge-dual of   $N_{\ha\hb\hc}$,  
$N_{\ha\hb}\equiv {1\over 6}\ve_{\ha\hb\hc\hd\he}N^{\hc\hd\he}$.
Then, the SU(2)-curvature can be rewritten in the form:
\bea
R_\hal^i{}_{\hbe}^j{}_{}{}^{kl}
&=&
{3\ri}\ve_{\hal\hbe}\ve^{ij}S^{kl}
+{\ri\over 2}(\G^\hc)_{\hal\hbe}\ve^{ij}T_{1\hc}{}^{kl}
+2\ri(\S^{\ha\hb})_{\hal\hbe}
\Big(F_{\ha\hb}+N_{\ha\hb}\Big)
\ve^{k(i}\ve^{j)l}
~.~~~~~~
\label{R-1-SU2-0}
\eea

Using the results obtained
and the fact that 
the constraint (\ref{constr-1}) is equivalent to  
\bea
T_{1\ha}{}^{jk}&=&0~,
\eea
eq. (\ref{Bianchi-1-2}) is now solved, 
and the dimension 1 torsion  becomes
\be
T_{\ha}{}_{\hbe}^j{}_{\hga}^{k}~=~
{1\over 2}(\Gamma_{\hat{a}})_{\hbe\hga}S^{jk}
+{1\over 2}\,\ve^{jk}(\Gamma^{\hat{b}})_{\hbe\hga}F_{\ha\hb}
-{1\over 4}\ve^{jk}(\Sigma^{\hb\hc})_{\hbe\hga}N_{\ha\hb\hc}
~.
\ee
The final form for the SU(2)-curvature is 
\bea
R_\hal^i{}_{\hbe}^j{}_{}{}^{kl}
&=&
{3\ri}\ve_{\hal\hbe}\ve^{ij}S^{kl}
+2\ri(\S^{\ha\hb})_{\hal\hbe}
\Big(F_{\ha\hb}+N_{\ha\hb}\Big)
\ve^{k(i}\ve^{j)l}
~.~~~~~~
\label{R-1-SU2}
\eea

\subsection{Solving the Bianchi identities: dimension 3/2}

${}$For dimension 3/2, the relevant Bianchi identities come
from both equations (\ref{Bianchi01}) and (\ref{Bianchi04}).
Setting ($\hat{A}=\ha,~\hat{B}=\hbeu,~\hat{C}=\hgau,~\hat{D}=\hdeu$)
in eq. (\ref{Bianchi01}) gives
\bea
0&=&\,
R_{\ha}{}_{\hbe}^j{}{}_{\,\hga}{}^{\hde}\d_{l}^{k}
+R_\ha{}_{\hga}^k{}_{\,\hbe}{}^{\hde}\d_{l}^{j}
+R_{\ha}{}_{\hbe}^j{}{}^k{}_l\d_{\hga}^{\hat{\d}}
+R_\ha{}_{\hga}^k{}{}^j{}_l\d_{\hbe}^{\hat{\d}}
\non\\
&&
+\cD_{\hbe}^jT_\ha{}_{\hga}^k{}^{\hde}_l
+2\ri \ve^{jk}(\Gamma^{\hat{e}})_{\hbe\hga}T_{\ha\hat{e}}{}^\hde_l
+\cD_{\hga}^kT_{\ha}{}_{\hbe}^j{}{}^{\hde}_l
~,\label{Bianchi-3/2-1-0}
\eea
while the choice
($\hat{A}=\ha,~\hat{B}=\hb,~\hat{C}=\hgau,~\hat{D}=\hd$) 
in (\ref{Bianchi01}) results in
\bea
0&=&R_{\hb}{}_{\hga}^k{}_\ha{}^\hd
-R_\ha{}_{\hga}^k{}_\hb{}^\hd
+2\ri T_{\ha\hb}{}^{\hrh k}(\G^\hd)_{\hrh\hga}
~. 
\label{Bianchi-3/2-2-0}
\eea
Choosing ($\hat{A}=\ha,~\hat{B}=\hb,~\hat{C}=\hgau$) in eq. (\ref{Bianchi04})
gives
\bea
0&=&2\ri T_{\ha\hb}{}_{\hga}^k
+\cD_\hga^k F_{\ha\hb}~.
\label{BianchiF13}
\eea

${}$For the analysis of the above identities, 
it is advantageous  to make use of the decomposition of 
a spin-tensor $A_{\ha\hb\hga}=-A_{\hb\ha\hga}$ 
into its irreducible components:
\begin{subequations}
\bea
A_{\ha\hb\hga}&=&\cA_{\ha\hb\hga}+2(\G_{{[}\ha})_\hga{}^\hde\cA_{\hb{]}\hde}+(\S_{\ha\hb})_\hga{}^\hde\cA_\hde~,
\quad
(\G^{\ha})_\hal{}^\hga\cA_{\ha\hga}\,=\,(\G^{\ha})_\hal{}^\hga\cA_{\ha\hb\hga}\,=\,0~.~~~~~
\label{decomposition_abga1}
\eea
\end{subequations}
Switching to the spinor notations, 
we have have to deal with 
\bea
A_{\hal\hbe\hga}:=\hf(\S^{\ha\hb})_{\hal\hbe}A_{\ha\hb\hga}=A_{(\hal\hbe)\hga}~,
\eea
and the corresponding decomposition is
\bea
&A_{\hal\hbe\hga}
=\cA_{\hal\hbe\hga}+\widetilde{\cA}_{\hga(\hal\hbe)}
+\ve_{\hga(\hal}\cA_{\hbe)}
~,~~~~~~\cA_{\hal\hbe\hga}=\cA_{(\hal\hbe\hga)}~,
\non
\\
&\widetilde{\cA}_{\hal\hbe\hga}:=(\G_{\ha})_{\hal\hbe}\cA^{\ha}{}_{\hga}
=-\widetilde{\cA}_{\hbe\hal\hga}
~,~~~
\ve^{\hal\hbe}\widetilde{\cA}_{\hal\hbe\hga}=\ve^{\hal\hga}\widetilde{\cA}_{\hal\hbe\hga}=0~,~~
\widetilde{\cA}_{[\hal\hbe\hga]}=0~.
~~~~~~
\label{decomposition_abga2}
\eea

${}$From equation (\ref{BianchiF13}) we immediately read off 
the dimension 3/2 torsion
\bea
T_{\ha\hb}{}^{\hga}_k&=&{\ri\over 2}\cD^\hga_k F_{\ha\hb}
~.
\label{BianchiF23}
\eea
Applying the decomposition (\ref{decomposition_abga2}) to the right-hand side 
of  (\ref{BianchiF23}) gives
\bea
\cD_{\hga}^kF_{\hal\hbe}&=&
 W_{\hal\hbe\hga}{}^{k}
+\X_{\hga(\hal\hbe)}{}^k
+\ve_{\hga(\hal}\cF_{\hbe)}{}^k~.
\label{decompositionFhalhbehga}
\eea
Next, equation (\ref{Bianchi-3/2-2-0}) is solved by 
\begin{subequations}
\bea
R_{\ha}{}_{\hbe}^j{}^{\hc\hd}&=&
\, \hf (\G_\ha)_{\hbe}{}^\hde\cD_\hde^{j}F^{\hc\hd}
-\hf (\G^\hc)_{\hbe}{}^\hde\cD_\hde^{j}F^\hd{}_{\ha}
+\hf (\G^\hd)_{\hbe}{}^\hde\cD_\hde^{j}F^\hc{}_{\ha}
~.
\label{Lorentz-3/2-0}
\eea
\end{subequations}

Equation (\ref{Bianchi-3/2-1-0}) 
allows us to compute the
SU(2)-curvature 
$R_{\ha}{}_{\hbe}^j{}{}^{kl}$.
Taking the trace over $\hga$ and $\hde$ in eq. (\ref{Bianchi-3/2-1-0}) 
gives
\bea
4R_{\ha}{}_{\hbe}^j{}{}^{kl}
+R_{\ha}{}_{\hbe}^k{}{}^{jl}
&=&\D_{\ha}{}_{\hbe}^j{}^{kl}~,
\label{Bianchi-3/2-1-3}
\eea
where
\bea
\D_{\ha}{}_{\hbe}^j{}^{kl}
&=&
-{7\over 4} (\G^\hb)_{\hbe}{}^\hga\cD_\hga^{k} F_{\ha\hb}\ve^{jl}
-\ve^{jk}(\Gamma^{\hat{b}})_{\hbe}{}^{\hga}\cD_{\hga}^{l} F_{\ha\hb}
-{1\over 8}\ve_{\ha\hb\hc\hd\he}(\S^{\hd\he})_{\hbe}{}^{\hga} \cD_{\hga}^{k}F^{\hb\hc}\ve^{jl}
\non\\
&&
+{1\over 4}\ve^{jl}(\Sigma^{\hb\hc})_{\hbe}{}^{\hga}\cD_{\hga}^kN_{\ha\hb\hc}
-{1\over 2}(\Gamma_{\hat{a}})_{\hbe}{}^{\hga}\cD_{\hga}^kS^{jl}~.
\eea
Equation (\ref{Bianchi-3/2-1-3}) is solved by
\bea
R_{\ha}{}_{\hbe}^j{}^{kl}&=&
{4\over 15}\D_{\ha}{}_{\hbe}^j{}^{kl}
-{1\over 15}\D_{\ha}{}_{\hbe}^k{}^{jl}~.
\label{R-ha-hal-su2-0}
\eea
Since $R_{\ha}{}_{\hbe}^j{}{}^{kl}$ is symmetric in $k$ and $l$,  
eq. (\ref{R-ha-hal-su2-0}) can be seen to be consistent under the  conditions:
\begin{subequations}
\bea
\cD_{\hal k}S^{kj}&=&
{3\over 20}(\S^{\ha\hb})_{\hal}{}^\hga\cD_\hga^{j} \Big(3F_{\ha\hb}+N_{\ha\hb}\Big)
~,\label{S-1}
\\
\cD_{\hga}^kN_{\hal\hbe}&=&
\cN_{\hal\hbe\hga}{}^{k}
+2\,\X_{\hga(\hal\hbe)}{}^k
+\ve_{\hga(\hal}\cN_{\hbe)}{}^k~.
\label{constr1-1}
\eea
\end{subequations}
Here $\X_{\ha\hbe}{}^k$
is the spin-vector  which occurs in (\ref{decompositionFhalhbehga}). 
At this point, the SU(2)-curvature 
has been completely determined.
\bea
R_{\ha}{}_{\hbe}^j{}^{kl}&=&
-{4\over 5} (\G^\hb)_{\hbe}{}^\hga\ve^{j(k}\cD_\hga^{l)} F_{\ha\hb}
-{1\over 30}\ve_{\ha\hb\hc\hd\he}(\S^{\hd\he})_{\hbe}{}^{\hga}\ve^{j(k}\cD_{\hga}^{l)}
\Big(F^{\hb\hc}+N^{\hb\hc}\Big)
\non\\
&&
-{2\over 15}(\Gamma_{\hat{a}})_{\hbe}{}^{\hga}\cD_{\hga}^{(k}S^{l)j}
+{1\over 30}(\Gamma_{\hat{a}})_{\hbe}{}^{\hga}\cD_{\hga}^jS^{kl}~.
\label{R-SU2-3/2-1}
\eea

Using the previous results, one can prove that 
equation (\ref{Bianchi-3/2-1-0}) implies the last two constraints:
\begin{subequations}
\bea
\cN_{\hal\hbe\hga}{}^k=-W_{\hal\hbe\hga}{}^k~,
~~~~~~~~~~~~~~~~~~~~~~~~~~~~~~
\label{Bianchi2N2}
\\
\cD_{\hbe}^{k}S^{jl}=
{1\over 10}(\S_{\ha\hb})_\hbe{}^\hde
\ve^{k(j}\cD_{\hde}^{l)}
\Big(3F^{\ha\hb}+N^{\ha\hb}\Big)
=
-{1\over 2}\ve^{k(j}\Big(3\cF_\hbe{}^{l)}+\cN_{\hbe}{}^{l)}\Big)
~.
\label{S}
\eea
\end{subequations}
It is important to note that (\ref{S}) implies equation (\ref{S-1}).

Expression (\ref{R-SU2-3/2-1})
can actually be further simplified,
using eqs. (\ref{decoDF2}, \ref{decoDF1}), (\ref{N-or-Bianchi2}, \ref{N-or-Bianchi}) 
and (\ref{S-Bianchi}). 
The final expression for the 
SU(2)-curvature is
\bea
R_{\ha}{}_{\hbe}^j{}{}^{kl}&=&
-3\ve^{j(k}\X_{\ha\hbe}{}^{l)}
+{5\over 4} (\G_{\ha})_{\hbe}{}^{\hal}\ve^{j(k}\cF_{\hal}{}^{l)}
-{1\over 4}(\G_\ha)_{\hbe}{}^{\hal}\ve^{j(k}\cN_\hal{}^{l)}~.
\eea

\subsection{Solving the Bianchi identities: dimension 2}

${}$For dimension 2, the relevant Bianchi identities 
 are generated 
from eq. (\ref{Bianchi01}) with ($\hat{A}=\ha,~\hat{B}=\hb,~\hat{C}=\hgau,~\hat{D}=\hdeu$)
\be
0=
R_{\ha\hb}{}^k{}_l\d_\hga^\hde
+R_{\ha\hb}{}_{\hga}{}^{\hde}\d^k_l
-\cD_\ha T_{\hb}{}_{\hga}^k{}^\hde_l
+\cD_\hb T_\ha{}_{\hga}^k{}^\hde_l
-T_{\hb}{}_{\hga}^k{}^\hrh_q\, T_\ha{}_{\hrh}^q{}^\hde_l
-\cD_\hga^k T_{\ha\hb}{}^\hde_l
+T_\ha{}_{\hga}^k{}^\hrh_q\, T_\hb{}_{\hrh}^q{}^\hde_l
~,
\label{Bianchi-2-1}
\ee
from (\ref{Bianchi01}) with ($\hat{A}=\ha,~\hat{B}=\hb,~\hat{C}=\hc,~\hat{D}=\hd$) 
\bea
0&=&R_{\ha\hb}{}_\hc{}^\hd
+R_{\hb\hc}{}_\ha{}^\hd
+R_{\hc\ha}{}_\hb{}^\hd
~,
\label{Bianchi2-2-0}
\eea
and also from (\ref{Bianchi04}) with ($\hat{A}=\ha,~\hat{B}=\hb,~\hat{C}=\hc$)
\bea
0&=&\cD_\ha F_{\hb\hc}
+\cD_\hb F_{\hc\ha}
+\cD_\hc F_{\ha\hb}~.
\label{Bianchi-F-2}
\eea

Let us first analyze eq. (\ref{Bianchi-2-1}). 
This can be used to extract the curvatures. 
We start by rewriting (\ref{Bianchi-2-1}) in the form:
\bea
R_{\ha\hb}{}^{kl}\ve_{\hga\hde}
+R_{\ha\hb\hga\hde}\ve^{kl}&=&
\D_{\ha\hb}{}_{\hga}^k{}_\hde^l
~,
\label{Bianchi-2-1-1}
\eea
with
\bea
\D_{\ha\hb}{}_{\hga}^k{}_\hde^l&=&
-{\ri\over 2}\cD_\hga^k \cD_\hde^lF_{\ha\hb}
-\ve^{kl}(\G^\hc)_{\hga\hde}\cD_{{[}\ha}F_{\hb{]}\hc}
+{1\over 4}\ve^{kl}\ve_{\hm\hn\hd\he{[}\ha}(\S^{\hm\hn})_{\hga\hde}\cD_{\hb{]}}N^{\hd\he}
+(\G_{{[}\ha})_{\hga\hde}\cD_{\hb{]}}S^{kl}
\non\\
&&
+\ve^{kl}(\S^{\hc\hd})_{\hga\hde}F_{\ha\hc}F_{\hb\hd}
+{1\over 4}\ve^{kl}F^\hm{}_{{[}\ha}\ve_{\hb{]}\hm\hn\hd\he}(\G^\hn)_{\hga\hde}N^{\hd\he}
-{2}(\S^{\hc}{}_{{[}\ha})_{\hga\hde}F_{\hb{]}\hc}S^{kl}
\non\\
&&
-{1\over 2}(\S^{\hc}{}_{[\ha})_{\hga\hde}N_{\hb]\hd}N^{\hc\hd}
-{1\over 4}(\S^{\hc\hd})_{\hga\hde}N^{\ha\hc}N^{\hb\hd}
-{1\over 8}(\S_{\ha\hb})_{\hga\hde}N^{\hc\hd}N_{\hc\hd}
\non\\
&&
+{1\over 4}\ve_{\ha\hb\hc\hd\he}(\G^\hc)_{\hga\hde}N^{\hd\he}S^{kl}
+{1\over 2}(\S_{\ha\hb})_{\hga\hde}\ve^{kl}S^{ij}S_{ij}
~.\label{also2}
\eea
Considering  the part of  (\ref{Bianchi-2-1-1}) which is symmetric 
in $\hga$ and $\hde$ and also antisymmetric in $k$ and $l$, 
we read off the expression for the
Lorentz curvature 
$R_{\ha\hb}{}^{\hc\hd}=-\hf\ve_{kl}(\S^{\hc\hd})^{\hga\hde}\D_{\ha\hb}{}_{\hga}^k{}_{\hde}^l$.
The result is
\bea
R_{\ha\hb}{}^{\hc\hd}&=&
\,{1\over 4}\ve^{\hc\hd}{}_{\hm\hn{[}\ha}\cD_{\hb{]}}N^{\hm\hn}
+{\ri\over 8}(\S^{\hc\hd})^{\hga\hde}\cD_{\hga}^k \cD_{\hde k}F_{\ha\hb}
-{1\over 8}\d_\ha^{[\hc}\d^{\hd]}_{\hb}N^{\hm\hn}N_{\hm\hn}
\non\\
&&
+\hf\d^{[\hc}_{{[}\ha}N_{\hb{]}\hm}N^{\hd]\hm}
-{1\over 4}N_\ha{}^{[\hc} N_\hb{}^{\hd]}
-F_{\ha}{}^{[\hc}F_{\hb}{}^{\hd]}
+{1\over 2}\d_\ha^{[\hc}\d^{\hd]}_{\hb}S^{ij}S_{ij}~.
\label{R-Lorentz-2-0}
\eea
Next, isolating  the part of (\ref{Bianchi-2-1-1}) which is
proportional to $\ve_{\hga\hde}$  and  
symmetric in $k$, $l$, we can determine the SU(2)-curvature
$R_{\ha\hb}{}^{kl}=-{1\over 4}\ve^{\hga\hde}\D_{\ha\hb}{}_{\hga}^k{}_\hde^l$.
The result is
\bea
R_{\ha\hb}{}^{kl}&=&
-{\ri\over 8}\cD^{\hga (k} \cD_\hga^{l)}F_{\ha\hb}~.
\label{R-SU(2)}
\eea

Equation (\ref{Bianchi-2-1-1}) has allowed us to determine the curvatures.
However it still contains some nontrivial information.
Using the relations (\ref{Bianchi-F-2}), (\ref{R-Lorentz-2-0}) and (\ref{R-SU(2)}),
eq. (\ref{Bianchi-2-1-1}) 
can be seen to reduce to
\bea
0&=&
-{\ri\over 8}(\G_\hc)^{\hga\hde}\cD_\hga^{(k} \cD_\hde^{l)}F_{\ha\hb}
-\eta_{\hc{[}\ha}\cD_{\hb{]}}S^{kl}
-{1\over 4}\ve_{\ha\hb\hc\hd\he}N^{\hd\he}S^{kl}
~.
\label{constr-2-2}
\eea
This implies
\bea
\cD_{\ha}S^{kl}=
{\ri\over 16}(\G^\hb)^{\hga\hde}\cD_\hga^{(k} \cD_\hde^{l)}F_{\ha\hb}
=
-{3\ri\over 16}\cD^{\hga(k}\X_{\ha\hga}{}^{l)}
-{\ri\over 8}(\G_\ha)^{\hga\hde}\cD_\hga^{(k}\cF_\hde{}^{l)}~.
\label{constr-2-2-2}
\eea
Next, due to the identity
\bea
\cD_{\ha}&=&\,
{\ri\over 8}\ve_{ij}(\G_\ha)^{\hal\hbe}\cD_\hal^i\cD_\hbe^j
+{1\over 8}\ve_{\ha\hb\hc\hd\he}N^{\hb\hc}M^{\hd\he}~,
\label{D_aVia2spinors}
\eea
and the dimension-$3/2$ constraint (\ref{S-Bianchi})
that determines $\cD_\hal^iS^{kl}$,
it also holds
\bea
\cD_{\ha}S^{kl}&=&-{3\ri\over 16}(\G_\ha)^{\hal\hbe}\cD_{\hal}^{(k}\cF_\hbe{}^{l)}
-{\ri\over 16}(\G_\ha)^{\hal\hbe}\cD_{\hal}^{(k}\cN_\hbe{}^{l)}~.
\label{constr-2-2-2-2}
\eea
Now, requiring 
the compatibility  of the equations (\ref{constr-2-2-2}) and (\ref{constr-2-2-2-2}), 
we generate  the constraint
\bea
(\G_\ha)^{\hal\hbe}\cD_{\hal}^{(k}\cN_\hbe{}^{l)}&=&-(\G_\ha)^{\hal\hbe}\cD_{\hal}^{(k}\cF_\hbe{}^{l)}
+3\cD^{\hga(k}\X_{\ha\hga}{}^{l)}~.
\eea
This 
turns out to be equivalent to (\ref{dim2-1}),  since 
\bea
\cD^{\hal(k}\cN_\hal{}^{l)}={1\over 5}\{\cD_\hal^{(k},\cD_\hbe^{l)}\}N^{\hal\hbe}
={4\ri\over 5}M_{\hal\hbe}N^{\hal\hbe}=0~.
\eea
Similar considerations give
$\cD^{\hal(k}\cF_\hal{}^{l)}=0$.

Further analysis of equation (\ref{constr-2-2}) leads to another 
constraint, eq.  (\ref{dim2-4}).

The Bianchi identity (\ref{Bianchi-F-2})
is equivalent to
$\ve^{\ha\hb\hc\hd\he}\cD_\hc F_{\hd\he}=0$.  
The latter can be rewritten, with the aid of (\ref{D_aVia2spinors}), 
as follows:
\bea
0&=&\ve^{\ha\hb\hc\hd\he}\Big(\,\ri(\G_\hc)^{\hal\hbe}\cD^k_\hal\cD_{\hbe k}F_{\hd\he}
+\ve_\hc{}^{\hm\hn\hp\hq}N_{\hm\hn}M_{\hp\hq}F_{\hd\he}\Big)~,
\eea
which, using (\ref{decoDF2}), can be seen to be equivalent to equation (\ref{dim2-5}).

Now, consider the Bianchi identity (\ref{Bianchi2-2-0}).
Using the Lorentz curvature (\ref{R-Lorentz-2-0}), 
eq. (\ref{Bianchi2-2-0}) turns out to be equivalent to
\bea
\cD_{\ha}N_{\hb\hc}&=&
\,
{\ri\over 8}\ve_{\hb\hc\hm\hn\hp}(\S_{\ha}{}^\hm)^{\hga\hde}\cD_{\hga}^k \cD_{\hde k}F^{\hn\hp}
+{\ri\over 12}\eta_{\ha[\hb}
\ve_{\hc]\hm\hn\hp\hq}(\S^{\hm\hn})^{\hga\hde}\cD_{\hga}^k \cD_{\hde k}F^{\hp\hq}
\non\\
&&
-{1\over 12}\ve_{\hb\hc\hm\hn\hp}\big(4F_\ha{}^\hm F^{\hn\hp}
+N_\ha{}^\hm N^{\hn\hp} \big)
~.
\label{2-4i-3-I}
\eea
The latter can rewritten as 
\bea
&&\cD_{\ha}N_{\hb\hc}=
{\ri\over 16}\ve_{\ha\hb\hc\hd\he}\cD^{\hde k} W^{\hd\he}{}_{\hde k}
-{\ri\over 8}(\G_\ha)^{\hga\hde}\cD_{\hga}^k W_{\hb\hc\hde k}
+{\ri\over 2}(\S_{\hb\hc})^{\hga\hde}\cD_{\hga}^k \X_{\ha\hde k}
+{\ri\over 8}\ve_{\ha\hb\hc\hm\hn}(\G^\hm)^{\hga\hde}\cD_{\hga}^k \X^\hn{}_{\hde k}
\non\\
&&
+{\ri\over 8}\eta_{\ha[\hb}(\G_{\hc]})^{\hga\hde}\cD_{\hga}^k \cF_{\hde k}
+{\ri\over 8}\ve_{\ha\hb\hc\hd\he}(\S^{\hd\he})^{\hga\hde}\cD_{\hga}^k \cF_{\hrh k}
-{1\over 12}\ve_{\hb\hc\hm\hn\hp}\big(4F_\ha{}^\hm F^{\hn\hp}
+N_\ha{}^\hm N^{\hn\hp} \big)
~.
\label{D_aN_{bc}2}
\eea
On the other hand, one can compute $\cD_{\ha}N_{\hb\hc}$
by using (\ref{D_aVia2spinors}) and 
the dimension $3/2$ Bianchi identity 
(\ref{N-or-Bianchi2}). 
Then one gets
\bea
&&\cD_{\ha}N_{\hb\hc}=\,
-{\ri\over 8}(\G_\ha)^{\hal\hbe}\cD^k_\hal W_{\hb\hc\hbe k}
+{\ri\over 2}\eta_{\ha[\hb}\cD^{\hal k}\X_{\hc]\hal k}
-\ri(\S_{\ha[\hb})^{\hal\hbe}\cD^k_\hal\X_{\hc]\hbe k}
\non\\
&&~~~
-{\ri\over 16}\ve_{\ha\hb\hc\hd\he}(\S^{\hd\he})^{\hal\hbe}\cD^k_\hal\cN_{\hbe k}
+{\ri\over 8}\eta_{\ha[\hb}(\G_{\hc]})^{\hal\hbe}\cD^k_\hal\cN_{\hbe k}
-{1\over 2}\ve_{\hm\hn\hp\ha[\hb}N_{\hc]}{}^{\hm}N^{\hn\hp}
~.
\label{D_aN_{bc}1}
\eea
Requiring  the equivalence of (\ref{D_aN_{bc}2}) and (\ref{D_aN_{bc}1}) 
 and making use of 
(\ref{dim2-5}), one obtains the  constraints
(\ref{dim2-2}) and (\ref{dim2-3}).

We have solved all Bianchi identities of dimension 2. 
Using the relations obtained, we can still simplify some of the results. 
Making use of  (\ref{D_aN_{bc}2}) allows us to rewrite 
the Lorentz curvature (\ref{R-Lorentz-2-0}) in the form:
\bea
R_{\ha\hb}{}^{\hc\hd}&=&\,
{\ri\over 24}(\S_{\ha\hb})^{\hga\hde}\cD_{\hga}^k \cD_{\hde k}F^{\hc\hd}
+{\ri\over 12}(\S_{[\ha}{}^{[\hc})^{\hga\hde}\cD_{\hga}^k \cD_{\hde k}F_{\hb]}{}^{\hd]}
+{\ri\over 24}(\S^{\hc\hd})^{\hga\hde}\cD_{\hga}^k \cD_{\hde k}F_{\ha\hb}
\non\\
&&
-{1\over 3}F_{\ha\hb} F^{\hc\hd}
-{1\over 3}F_{\ha}{}^{[\hc} F_\hb{}^{\hd]}
-{1\over 12}N_{\ha\hb} N^{\hc\hd}
-{1\over 12}N_{\ha}{}^{[\hc} N_\hb{}^{\hd]}
+{1\over 2}\d^{[\hc}_{[\ha}N_{\hb]\hm}N^{\hd{]}\hm}
\non\\
&&
-{1\over 8}\d_\ha^{[\hc}\d^{\hd]}_{\hb}N^{\hm\hn}N_{\hm\hn}
+{1\over 2}\d_\ha^{[\hc}\d^{\hd]}_{\hb}S^{ij}S_{ij}~.
\label{LorentzCurvature2-2}
\eea
Next, using the equation
\bea
\cD^{\hga(k}W_{\ha\hb\hga}{}^{l)}&=&\,
3(\S_{\ha\hb})^{\hga\hde}\cD_\hga^{(k}\cF_\hde{}^{l)}
-4(\G_{[\ha})^{\hga\hde}\cD_\hga^{(k}\X_{\hb]\hde}{}^{l)}
+12\ri N_{\ha\hb}S^{kl}~,
\eea
which follows from (\ref{dim2-4}), one can see that the  SU(2)-curvature (\ref{R-SU(2)})
can be rewritten as follows:
\bea
R_{\ha\hb}{}^{kl}&=&
{3\ri\over 4}(\G_{[\ha})^{\hga\hde}\cD_\hga^{(k}\X_{\hb]\hde}{}^{l)}
-{\ri\over 4}(\S_{\ha\hb})^{\hga\hde}\cD_\hga^{(k}\cF_\hde{}^{l)}
+{3\over 2} N_{\ha\hb}S^{kl}
~.
\label{SU(2)Curvature2-2}
\eea

${}$Finally let us turn to the Bianchi identities of dimension 5/2 and 3.
${}$For dimension 5/2, there is only one nontrivial Bianchi identity. 
This is the identity 
(\ref{Bianchi01}) with ($\hat{A}=\ha,~\hat{B}=\hb,~\hat{C}=\hc,~\hat{D}=\hdeu$)
\bea
0&=&-\cD_\ha T_{\hb\hc}{}^\hde_l
+T_{\ha\hb}{}^{\hrh}_q\,T_\hc{}_{\hrh}^q{}^\hde_l
-\cD_\hb T_{\hc\ha}{}^\hde_l
-T_{\hb\hc}{}^{\hrh}_q\,T_{\ha}{}_{\hrh}^q{}^\hde_l
-\cD_\hc T_{\ha\hb}{}^\hde_l
-T_{\hc\ha}{}^{\hrh}_q\,T_\hb{}_{\hrh}^q{}^\hde_l~.
\label{Bianchi5/2-1}
\eea
This equation can be seen to be satisfied identically provided the Bianchi identities 
of lower dimension hold.
${}$For dimension 3, there are no nontrivial  Bianchi identities.

\sect{Projective superspace formalism}
\label{APS}

The projective superspace approach was originally formulated
for rigid supersymmetric theories with eight supercharges 
in four space-time dimensions \cite{KLR,LR1}, and later it was   
generalized to five \cite{KuzLin} and six \cite{GL,GPT} dimensions. 
Superconformal field theory in projective superspace has also been developed in four 
and five dimensions \cite{K,K2}.

As demonstrated in \cite{KT-Msugra5D},  the concept of projective supermultiplets
can naturally be extended to the case of 5D $\cN=1$ supergravity.
In this section,  we first recall the definition of covariant projective multiplets in curved superspace, 
following \cite{KT-Msugra5D}. After that we formulate a manifestly locally supersymmetric 
action principle. 

To start with, it is instructive to recall 
the kinematical setup for projective superspace 
in the case of 5D $\cN=1$ supersymmetry. Let ${\mathbb R}^{5|8}$ denote
the flat global superspace  parametrized by  coordinates 
 $z^{\hat{A}}=(x^{\ha},\q^{\hat{\a}}_i)$. The corresponding covariant derivatives
 $D_{\hat{A}} =(\pa_{\hat{a}}, D_{\hat{\a}}^i) $ obey the algebra
\be
\{D^i_{\hat \a} \, , \, D^j_{\hat \b} \} = -2{\rm i} \,
\ve^{ij}\,
\Big( (\G^{\hat c} ){}_{\hat \a \hat \b} \, \pa_{\hat c} 
+ \ve_{\hat \a \hat \b} \,Z \Big)~,
\qquad [ D^i_{\hat \a} \, , \, \pa_{\hat b} ] = 
[ D^i_{\hat \a} \, , \, Z ] =0~, 
\ee
which follows from (\ref{covDev2spinor-})--(\ref{covDev2spinor-3})
by setting $S^{ij}= N_{\hat a \hat b} = F_{\hat a \hat b}=0$.
Making use of  an isotwistor $u^+_i \in {\mathbb C}^2 \setminus \{0\}$
allows one to introduce 
a subset of strictly anti-commuting  spinor covariant derivatives
$D^+_{\hat \a} :=u^+_i D^i_{\hat \a} $.
\bea
\{ D^+_{\hat \a} , D^+_{\hat \b} \} =0~.
\label{flatconstr}
\eea
Hence, one can define so-called {\it analytic} superfields 
$Q(z, u^+)$ constrained by 
$D^+_{\hat \a} Q  =0$.
Such a superfield $ Q(z, u^+)$ is called a {\it projective supermultiplet}, if
it is holomorphic (on an open subset of ${\mathbb C}^2 \setminus \{0\}$)
and a homogeneous function of $u^+$, 
$Q(z, c\, u^+) =c^n \, Q(z, u^+)$, with $c\in {\mathbb C}^*$. 
The isotwistor
$u^+_i \in {\mathbb C}^2 \setminus\{0\}$ appears to be  defined modulo the equivalence relation
$u^+_i \sim c\,u^+_i$,  with $c\in {\mathbb C}^*$, 
since this is true for both 
the  constraint $D^+_{\hat \a} Q  =0$ and the superfield $ Q(z,u^+)$ itself.
As a result, the projective multiplets live in
the  projective superspace  ${\mathbb R}^{4|8} \times {\mathbb C}P^1$.

\subsection{Projective supermultiplets}

In curved superspace, the isotwistor  variables $u^{+}_i \in  
{\mathbb C}^2 \setminus  \{0\}$ are defined to be inert with respect to 
the local group SU(2)  \cite{KT-Msugra5D} (see also \cite{KT-M}).
Instead of the anticommutation relation (\ref{flatconstr}),
the operators $\cD^+_{\hat \a}:=u^+_i\,\cD^i_{\hat \a} $  obey the following algebra:
\bea
\{  \cD^+_{\hat \a} , \cD^+_{\hat \b} \}
=-4{\rm i}\, \Big(F_{\hal \hbe}+N_{\hal \hbe}\Big)\,J^{++}
+4{\rm i} \, S^{++}M_{\hal \hbe}~,
\label{analyt}
\eea
where 
$J^{++}:=u^+_i u^+_j \,J^{ij}$ and 
$S^{++}:=u^+_i u^+_j \,S^{ij}$. 
Eq. (\ref{analyt}) follows from (\ref{covDev2spinor-}). 
Now, for the constraint $\cD^+_{\hat \a} Q =0$ 
to be consistent,  $Q(z,u^+)$  must be scalar with respect to the Lorentz group,
$ M_{\hal \hbe} Q=0$, and also possess special properties with respect to the
group SU(2), that is,  $J^{++}Q=0$. 
Let us  define such multiplets,  following \cite{KT-Msugra5D}.

A projective supermultiplet of weight $n$,
$Q^{(n)}(z,u^+)$, is a scalar superfield that 
lives on  $\cM^{5|8}$, 
is holomorphic with respect to 
the isotwistor variables $u^{+}_i $ on an open domain of 
${\mathbb C}^2 \setminus  \{0\}$, 
and is characterized by the following conditions:\\
(i) it obeys the covariant analyticity constraint 
\be
\cD^+_{\hat \a} Q^{(n)}  =0~;
\label{ana}
\ee  
(ii) it is  a homogeneous function of $u^+$ 
of degree $n$, that is,  
\be
Q^{(n)}(z,c\,u^+)\,=\,c^n\,Q^{(n)}(z,u^+)~, \qquad c\in \mathbb{C}^*~;
\label{weight}
\ee
(iii) infinitesimal gauge transformations (\ref{tau}) act on $Q^{(n)}$ 
as follows:
\bea
\d Q^{(n)} 
&=& \Big( K^{\hat{C}} \cD_{\hat{C}} + K^{ij} J_{ij} \Big)Q^{(n)} ~,  
\non \\ 
K^{ij} J_{ij}  Q^{(n)}&=& -\frac{1}{(u^+u^-)} \Big(K^{++} D^{--} 
-n \, K^{+-}\Big) Q^{(n)} ~, \qquad 
K^{\pm \pm } =K^{ij}\, u^{\pm}_i u^{\pm}_j ~,
\label{harmult1}   
\eea 
where
\bea
D^{--}=u^{-i}\frac{\partial}{\partial u^{+ i}} ~,\qquad
D^{++}=u^{+ i}\frac{\partial}{\partial u^{- i}} ~.
\label{5}
\eea
The transformation law (\ref{harmult1}) involves an additional isotwistor,  $u^-_i$, 
which is subject 
to the only condition $(u^+u^-) = u^{+i}u^-_i \neq 0$, and is otherwise completely arbitrary.
By construction, $Q^{(n)}$ is independent of $u^-$, 
i.e. $\pa  Q^{(n)} / \pa u^{-i} =0$,
and hence $D^{++}Q^{(n)}=0$.
One can see that $\d Q^{(n)} $ 
is also independent of the isotwistor $u^-$, $\pa (\d Q^{(n)})/\pa u^{-i} =0$,
due to (\ref{weight}). 
It follows  from (\ref{harmult1})
\bea
J^{++} \,Q^{(n)}=0~, \qquad J^{++} \propto D^{++}~,
\label{J++}
\eea
and hence the covariant analyticity constraint (\ref{ana}) is indeed consistent.

The transformation law (\ref{harmult1}) is a generalization of that for superconformal 
projective supermultiplets in four and five dimensions \cite{K,K2} 
and for projective supermultiplets in the 5D $\cN=1$ anti-de Sitter superspace 
\cite{KT-M}.

It should be pointed out that the  transformation law (\ref{harmult1}) corresponds
to the projective supermultiplets with zero central charge, $ZQ^{(n)}=0$. 
Such off-shell multiplets are most interesting for applications, and our consideration 
will be restricted to their study. It is not difficult, however, to modify    (\ref{harmult1})
in order to be applicable to the case of  off-shell   projective supermultiplets 
with an intrinsic zero central charge. 
The corresponding transformation law is  \cite{KT-Msugra5D}
\bea
\d Q^{(n)} 
&=& \Big( K^{\hat{C}} \cD_{\hat{C}} + K^{ij} J_{ij} 
+\t Z \Big)Q^{(n)} ~.
\eea
As an example, we can consider an off-shell hypermultiplet with intrinsic central charge, 
which is described by $q^+(z,u^+) = u^+_i q^i(z) $. 
It is this realization\footnote{This is a generalization of the Sohnius off-shell formulation 
for  hypermultiplet \cite{Sohnius}.}
which is used in the component approaches of \cite{Zucker,Ohashi}.
In this realization, the hypermultiplet  becomes on-shell provided $Zq^{+}=0$. 

Given a projective multiplet $Q^{(n)}$,
its complex conjugate 
is not covariantly analytic.
However, similar to the flat four-dimensional case  
\cite{GIKOS,KLR}  (see also \cite{KT-M}),
one can introduce a generalized,  analyticity-preserving 
conjugation, $Q^{(n)} \to \widetilde{Q}^{(n)}$, defined as
\be
\widetilde{Q}^{(n)} (u^+)\equiv \bar{Q}^{(n)}\big(
\overline{u^+} \to 
\widetilde{u}^+\big)~, 
\qquad \widetilde{u}^+ = {\rm i}\, \s_2\, u^+~, 
\ee
with $\bar{Q}^{(n)}(\overline{u^+}) $ the complex conjugate of $Q^{(n)}$.
Its fundamental property is
\bea
\widetilde{ {\cD^+_{\hat \a} Q^{(n)}} }=(-1)^{\e(Q^{(n)})}\, \cD^{+\hat \a}
 \widetilde{Q}{}^{(n)}~. 
\eea
One can see that
$\widetilde{\widetilde{Q}}{}^{(n)}=(-1)^nQ^{(n)}$,
and therefore real supermultiplets can be consistently defined when 
$n$ is even.
In what follows, $\widetilde{Q}^{(n)}$ will be called the smile-conjugate of 
${Q}^{(n)}$.

Examples of projective supermultiplets are given in \cite{KT-Msugra5D}, 
and the interested reader is referred to that paper for more details.

It follows from (\ref{S-Bianchi}) that $S^{++}$ is a projective superfield of weight two, 
\be
\cD^+_{\hat \a} S^{++}  =0~.
\ee

\subsection{Locally supersymmetric action}

Let $\cL^{++}$ be a real projective multiplet of weight two. 
Associated with $\cL^{++}$  is the following functional 
\bea
S(\cL^{++})&=&
\frac{1}{6\pi} \oint (u^+ \rd u^{+})
\int \rd^5 x \,{\rm d}^8\q\,E\, \frac{\cL^{++}}{(S^{++})^2}~, 
\qquad E^{-1}={\rm Ber}\, (E_{\hat A}{}^{\hat M})~.
\label{InvarAc}
\eea
We are going to show that $S$ defines a locally supersymmetric action principle.
This functional is obviously invariant under projective re-scalings
$u_i^+\to c\,u^+_i$.
Moreover, it turns out to be invariant under 
infinitesimal  gauge transformations (\ref{tau}) and (\ref{harmult1}). 
To prove the invariance under arbitrary supergravity gauge transformations, 
we first point out that  $Q^{(-2)}:= \cL^{++} /(S^{++})^2$ is a projective multiplet of weight $-2$,
since both $\cL^{++}$ and $S^{++}$ are projective multiplet of weight $+2$.
${}$For  $Q^{(-2)}$ the second line in (\ref{harmult1}) implies
\bea
K^{ij} J_{ij} \, Q^{(-2)}= -\frac{1}{(u^+u^-)} D^{--}\Big( K^{++} Q^{(-2)}\Big)~.
\eea
Next, since $K^{++} Q^{(-2)}$ has weight zero, 
it is easy to see 
\bea
(u^+ \rd u^{+} )\, K^{ij} J_{ij}  \,Q^{(-2)} = -{\rm d}t \,
\frac{{ \rm d}  }{{\rm d}t} \, Q^{(-2)}~, 
\eea
with $t$ the evolution parameter along the integration contour in 
(\ref{InvarAc}). Since the integration contour is closed, the SU(2)-part of 
the transformation (\ref{harmult1}) does not contribute to the variation of 
the action (\ref{InvarAc}). To complete the proof, it remains to take into the account 
the fact that $Q^{(-2)}$ is a Lorentz scalar.

Introduce the following fourth-order operator\footnote{This operator 
was  introduced in the case of 5D $\cN=1$ anti-de Sitter supersymmetry 
in \cite{KT-M}.}
\bea
\D^{(+4)} = ({\cD}^+)^4 -\frac{5}{12} {\rm i}\, S^{++}\,({\cD}^+)^2
+3 (S^{++})^2~,
\label{anapro}
\eea
where 
\bea
(\cD^+)^4:=-{1\over 96}\ve^{\hal\hbe\hga\hde}
\cD^+_{\hal} \cD^+_{\hbe}\cD^+_{\hga}\cD^+_{\hde}~, \qquad
(\cD^+)^2:=\cD^{+\hal}\cD^+_{\hal}~.
\eea
Its crucial property is that the superfield $Q^{(n)} $ defined by
\bea
Q^{(n)} (z,u^+) := \D^{(+4)} U^{(n-4)} (z,u^+) ~, 
\eea
is a weight-$n$ projective multiplet, 
\bea
\cD^+_{\hat \a}Q^{(n)}=0~,
\eea
for any  {\it unconstrained} scalar superfield  $U^{(n-4)} (z,u^+) $  that
lives on  $\cM^{5|8}$, 
is holomorphic with respect to 
the isotwistor variables $u^{+}_i $ on an open domain of 
${\mathbb C}^2 \setminus  \{0\}$, 
and is characterized by the following conditions:\\
(i) it is  a homogeneous function of $u^+$ 
of degree $n-4$, that is,  
\be
U^{(n-4)}(z,c\,u^+)\,=\,c^{n-4}\,U^{(n-4)}(z,u^+)~, \qquad c\in \mathbb{C}^*~;
\ee
(iii) infinitesimal gauge transformations (\ref{tau}) act on $U^{(n-4)}$ 
as follows:
\bea
\d U^{(n-4)} 
&=& \Big( K^{\hat{C}} \cD_{\hat{C}} + K^{ij} J_{ij}  \Big)U^{(n-4)} ~,  
\non \\ 
K^{ij} J_{ij}  \,U^{(n-4)}&=& -\frac{1}{(u^+u^-)} \Big(K^{++} D^{--} 
-(n-4) \, K^{+-}\Big) U^{(n-4)} ~. \eea 
We will call $U^{(n-4)}(z,c\,u^+)$ a {\it projective prepotential} 
for $Q^{(n)} $. 

The fourth-order operator (\ref{anapro}) is analogous to 
 the chiral projector in 4D $\cN=1$ supergravity \cite{Zumino}.

Let $U^{(-2)}$ be a projective prepotential for the Lagrangian $\cL^{++}$ 
in (\ref{InvarAc}). Representing 
\bea
U^{(-2)} = \frac{1}{3(S^{++})^2} \Big\{ \cL^{++} -
 ({\cD}^+)^4 U^{(-2)}  + \frac{5}{12} {\rm i}\, S^{++}\,({\cD}^+)^2 U^{(-2)} \Big\}~,
\label{U-->L}
\eea
we obtain
\bea
\frac{1}{2\pi} \oint (u^+\rd u^{+})
\int \rd^5 x \,{\rm d}^8\q\,E\, U^{(-2)} = 
\frac{1}{6\pi} 
\oint (u^+\rd u^{+})
\int \rd^5 x \,{\rm d}^8\q\,E\, \frac{\cL^{++}}{(S^{++})^2}~.
\label{InvarAc2}
\eea
One can see that the derivative terms in (\ref{U-->L}) do not contribute
to the integral in (\ref{InvarAc2}), as a consequence of the
anti-commutation relations (\ref{covDev2spinor-})--(\ref{covDev2spinor-3}).

Our action (\ref{InvarAc}) can be compared with the chiral action 
in 4D $\cN=1$ supergravity \cite{Zumino,SG}
(see also \cite{GGRS,BK} for reviews).

In the case of flat superspace, one  can not make use of (\ref{InvarAc2}).
Instead, here one can apply  the following relations
\bea
\frac{1}{2\pi} \oint (u^+\rd u^{+})
\int \rd^5 x \,{\rm d}^8\q\, U^{(-2)} &=& 
\frac{1}{2\pi} \oint \frac{(u^+\rd u^{+})}{(u^+u^-)^4}
\int \rd^5 x \, (D^-)^4(D^+)^4U^{(-2)} \Big|_{\q=0} \non \\
&=& \frac{1}{2\pi}  \oint \frac{(u^+\rd u^{+})}{(u^+u^-)^4}
\int \rd^5 x \, (D^-)^4 L^{++}\Big|_{\q=0}~,
\label{flatac}
\eea
with  $L^{++} := (D^+)^4U^{(-2)}$ the flat-superspace Lagrangan. 
Here 
\bea
(D^-)^4:=-{1\over 96}\ve^{\hal\hbe\hga\hde}
D^-_{\hal} D^-_{\hbe}D^-_{\hga}D^-_{\hde}~, \qquad
D^-_{\hat \a} := u^-_i D^i_{\hat \a}~.
\eea
The expression in the second line of (\ref{flatac}) is the rigid supersymmetric  action 
in 5D $\cN=1$ projective superspace \cite{K}. The latter is a natural generalization of the 
4D $\cN=2$ projective-superspace action originally given in \cite{KLR} and reformulated in 
a projective-invariant form in \cite{S}.
This action can be seen to be invariant 
under arbitrary transformations of the form:
\be
(u_i{}^-\,,\,u_i{}^+)~\to~(u_i{}^-\,,\, u_i{}^+ )\,R~,~~~~~~R\,=\,
\left(\begin{array}{cc}a~&0\\ b~&c~\end{array}\right)\,\in\,{\rm GL(2,\mathbb{C})}~.
\label{projectiveGaugeVar}
\ee
The same invariance  obviously holds for  the curved-superspace action 
(\ref{InvarAc}), for it is explicitly   independent of $u^-$.

Projective invariance (\ref{projectiveGaugeVar}) is an obvious property 
of the manifestly locally supersymmetric action (\ref{InvarAc}). 
As shown in section 5, it becomes a powerful constructive principle
when one is interested in reducing  the action to components in the Wess-Zumino gauge.

\sect{Wess-Zumino gauge}
\label{WZgauge}

In this section we elaborate the Wess-Zumino gauge 
for the 5D minimal supergravity multiplet, which was used in \cite{KT-Msugra5D}.
Our consideration will be similar to that originally given,  many years ago, for 
4D $\cN=1$ supergravity \cite{Zumino,WZ,RL} 
and then presented in a universally applicable form in  \cite{GGRS}.

Given a superfield $U(z)=U(x,\q)$, it is standard to denote 
as $U|$ its  $\q$-independent component, 
$U|:= U(x,\q=0)$.
 The Wess-Zumino (WZ) gauge 
for 5D $\cN=1$ supergravity  is defined by 
\bea
\cD_\ha\big|&=&\CD_\ha+\J_\ha{}^\hga_k(x) \cD^k_\hga\big|+\f_\ha{}^{kl}(x) J_{kl}+\cV_\ha(x) Z~,
\qquad
\cD^i_{\hat \a} \big|= \frac{\pa}{\pa \q^{\hat \a}_i}~.
\label{WZgauge1-1}
\eea
Here $\nabla_{\hat a}$ denotes the space-time covariant derivatives, 
\be 
\nabla_{\hat a} = e_{\hat{a}} + \o_{\hat{a}} ~, \qquad 
e_{\hat a} = e_{\hat a}{}^{\hat m} (x)\, \pa_{\hat m}~, 
\qquad  
 \o_{\hat{a}} =\hf \, \o_{\hat{a}}{}^{\hb \hc} (x) \,M_{\hb \hc}
 = \o_{\hat{a}}{}^{\hbe\hga} (x) \,M_{\hbe\hga}~,
\ee
with $e_{\hat a}{}^{\hat m} $ the component inverse vielbein, 
and $\o_{\hat{a}}{}^{\hb \hc}$ the Lorentz connection. 
The operators $\nabla_{\hat a}$ obey
 commutation relations of the form
\bea
\big[ \nabla_{\hat a} , \nabla_{\hat b} \big] = 
{\cT}_{\ha\hb}{}^{\hc} (x) \, \nabla_{\hat c} 
+\hf \cR_{\hat{a}\hat{b}}{}^{\hat{c}\hat{d}}(x)M_{\hat{c}\hat{d}}~,
\label{WZ2spD1}
\eea
with ${\cT}_{\ha\hb}{}^{\hc} $ the torsion, and $\cR_{\hat{a}\hat{b}}{}^{\hat{c}\hat{d}}$
the curvature.
Next, $ \J_\ha{}^\hga_k$ is the component gravitino,
while $\f_\ha{}^{kl} = \F_\ha{}^{kl}|$ and $\cV_{\hat a} = V_{\hat a}|$ 
are the component SU(2) and central-charge gauge fields, respectively. 
In addition to the geometric fields present in  (\ref{WZgauge1-1}), 
the supergravity multiplet includes some additional 
component fields which can be chosen as follows: $S_{ij}\big|$, $N_{\hat a \hat b} \big|$, 
$\cD^j_{\hat \a} S_{ij}\big|$ and $\cD^{\hat \a i} \cD^j_{\hat \a} S_{ij}\big|$.
All these fields, which  survive in the WZ gauge, 
constitute the 5D minimal supergravity multiplet  \cite{Howe5Dsugra}.

Making use of (\ref{WZgauge1-1})
one can readily obtain
\bea
{[}\cD_\ha,\cD_\hb{]}\big|&=&\,
 {[}\CD_\ha,\CD_\hb{]}
-2\J_{[\ha}{}^\hga_k\, {[}\cD_{\hb]},\cD_\hga^k{]}\big|
+\J_{\ha}{}^\hga_k\J_{\hb}{}^\hde_l \, \{\cD_\hga^k,\cD_\hde^l\}\big|
+2(\CD_{[\ha}\cV_{\hb]})Z
\non\\
&&
+2\Big(\CD_{[\ha}\J_{\hb]}{}^\hga_k-\phi_{[\ha}{}_k{}^l\J_{\hb]}{}^\hga_l\Big)\cD^k_\hga\big|
+2\Big(\CD_{[\ha}\phi_{\hb]}{}^{kl}+\phi_{[\ha}{}^{k}{}_j\phi_{\hb]}{}^{jl}\Big)J_{kl}
~.
\label{WZ2spD2}
\eea
This relation  can be simplified considerably by evaluating 
the (anti-)commutators ${[}\cD_\ha,\cD_\hb{]}$, 
${[}\cD_{\hb},\cD_\hga^k{]}$ and $\{\cD_\hga^k,\cD_\hde^l\}$ 
with the aid of  (\ref{covDev2spinor-})--(\ref{covDev2spinor-3}).
As a result, eq. (\ref{WZ2spD2}) can be seen to be equivalent to 
the following relations:
\begin{subequations}
\bea
{\cT}_{\ha\hb}{}^{\hc}&=&-2\ri\J_\ha{}^{\hga k}(\G^\hc)_{\hga\hde}\J_\hb{}^\hde_k~,
\label{WZ11}\\
T_{\ha\hb}{}^\hga_k\big|={\ri\over 2}\cD^\hga_k F_{\ha\hb}\big|&=&
2\CD_{[\ha}\J_{\hb]}{}^\hga_k
-2\phi_{[\ha}{}_k{}^j\J_{\hb]}{}^\hga_j
-{\cT}_{\ha\hb}{}^{\hc}\J_{\hc}{}^\hga_k
-2\J_{[\ha}{}^\hbe_jT_{\hb]}{}_{\hbe}^j\,{}^\hga_k\big|
~,
\label{WZ12}
\eea
\end{subequations}
as well as
\begin{subequations}
\bea
R_{\ha\hb}{}^{\hc\hd}\big|
&=&
\cR_{\hat{a}\hat{b}}{}^{\hat{c}\hat{d}}
-2\J_{[\ha}{}^\hga_k R_{\hb]}{}_{\hga}^k\,{}^{\hc\hd}\big|
+\J_{\ha}{}^\hga_k\J_{\hb}{}^\hde_lR_{\hga}^k{}_{\hde}^l\,{}^{\hc\hd}\big|~,
\label{curvature-Lor}
\\
F_{\ha\hb}\big|&=&\,
2(\CD_{[\ha}\cV_{\hb]})
+2\ri\J_{\ha}{}^{\hga k}\J_{\hb}{}^\hde_k\cV_{\hga\hde} 
-2\ri\J_{\ha}{}^\hga_k\J_{\hb}{}_\hga^k~,
\label{curvature-cc}
\\
R_{\ha\hb}{}^{ij}\big|
&=&\, 2\CD_{[\ha}\phi_{\hb]}{}^{ij}
+2\phi_{[\ha}{}^{i}{}_k\phi_{\hb ]}{}^{kj} \non \\
&& -2\J_{[\ha}{}^\hga_kR_{\hb]}{}_{\hga}^k\,{}^{ij}\big|
+\J_{\ha}{}^\hga_k\J_{\hb}{}^\hde_lR_{\hga}^k{}_{\hde}^l\,{}^{ij}\big|
+2\ri\J_{\ha}{}^{\hga k}\J_{\hb}{}^\hde_k\f_{\hga\hde}{}^{ij}
~.~~~~~~
\label{curvature-su(2)}
\eea
\end{subequations}
Eq.  (\ref{WZ11}) determines the space-time torsion in terms of the gravitino.
Eq. (\ref{WZ12}) constitutes a locally supersymmetric  version of the gravitino field
strength. Finally, 
eqs. (\ref{curvature-Lor})--(\ref{curvature-su(2)}) express the leading components
of the superspace curvature tensors in terms of the component fields.
Equations (\ref{WZ11}) and (\ref{WZ12}) will frequently be used  in section \ref{WZaction}.
The space-time torsion (\ref{WZ11}) will be especially important for the considerations
in section \ref{WZaction}, for it occurs in the rule 
for integration by parts:
\bea
\int \rd^5x\, e\,\CD_\ha U^\ha&=&\int \rd^5x\, e\,{\cT}_{\ha\hb}{}^{\hb}
\,U^\ha~, \qquad e^{-1} = \det \big(e_{\hat a}{}^{\hat m} \big)~.
\label{intPart}
\eea

In the WZ gauge, the supergravity gauge fredom (\ref{tau}) 
reduces to those transformations which preserve the WZ gauge. 
This is equivalent to the requirement 
\bea
0&=&\d\cD_\hal^i\big|=-\Big{[}K^\hbe_j\cD_\hbe^j+K^\hb\cD_\hb+K^{\hbe\hga}M_{\hbe\hga}
+K^{jk}J_{jk}+\tau Z,\cD_\hal^i\Big{]}\Big|~.
\eea
It implies the following restrictions on the transformation parameters:
\bea
\cD^i_{\hat \a} K^{\hat \b}_j\big| &=& K^{\hat c}\big| \,T_{\hat c}{ \,}^i_{\hat \a  }{}^{\hat \b}_j \big|
+ \d^i_j \, K_{\hat \a}{}^{\hat \b}\big| +\d_{\hat \a}^{\hat \b} K^i{}_j \big|~, \qquad 
\cD^i_{\hat \a} K^{\hat b}\big| = 
-2{\rm i}\, (\G^{\hat b})_{\hat \a \hat \g} \,K^{\hat \g i}\big|~,
\non \\
\cD^i_{\hat \a} K^{\hat \b \hat \g}\big| &=& K^{\hat C}\big| \,
R_{\hat C}{ }^i_{\hat \a  }{}^{\hat \b \hat \g} \big|~, \qquad 
\cD^i_{\hat \a} K^{jk}\big| = K^{\hat C}\big| \,
R_{\hat C}{ }^i_{\hat \a  }{}^{jk} \big|~, \qquad 
\cD^i_{\hat \a} \t \big| =-2{\rm i}\, K^i_{\hat \a}\big|~.
\eea

In the WZ gauge,  the transformation laws of the  gauge fields 
can be derived from 
\bea
\d\cD_\ha\big|&=&\d\CD_\ha+\d\J_\ha{}_j^\hbe\cD^j_\hbe\big|+\d\phi_\ha{}^{kl}J_{kl}+\d \cV_\ha Z
\non\\
&=&
-\Big{[}K^{\hat{B}}\cD_{\hat{B}}+K^{\hbe\hga}M_{\hbe\hga}
+K^{jk}J_{jk}+\tau Z,\cD_\ha\Big{]}\Big|~.
\eea
Some computations lead to
\begin{subequations}
\bea
\d e_\ha{}^\hm&=&\Big(\CD_\ha K^\hb\big|-2\ri\J_{\ha}{}^\hal_k(\G^\hb)_{\hal\hbe}K^{\hbe k}\big|
-K_\ha{}^\hb\big|\Big)e_\hb{}^\hm~,
\\
\d\o_\ha{}^{\hbe\hga}&=&
\,\Big(\CD_\ha K^\hb\big|-2\ri\J_{\ha}{}^\hal_k(\G^\hb)_{\hal\hde}K^{\hde k}\big|
-K_\ha{}^\hb\big|\Big)\o_\hb{}^{\hbe\hga}
+\CD_\ha K^{\hga\hde}\big|
\non\\
&&
+\J_{\ha}{}^\hbe_jK^{\hat{C}}\big|R_{\hat{C}}{}_\hbe^j{}^{\hga\hde}\big|
-K^{\hat{B}}\big|R_{\hat{B}\ha}{}^{\hga\hde}\big|~,
\eea 
\bea
\d\J_{\ha}{}^\hbe_j&=&\CD_\ha K^\hbe_j\big|
-\phi_{\ha}{}_j{}^kK_k^\hbe\big|
-2\ri\J_{\ha}{}_j^\hga(\G^\hc)_{\hga\hde}\J_{\hc}{}^\hbe_k K^{\hde k}\big|
+K^\hga_k\big|T_{\ha}{}^k_\hga{}^\hbe_j\big|
+\CD_\ha K^\hc\big|\J_{\hc}{}^\hbe_j
\non\\
&&
+\J_{\ha}{}^\hga_kK^\hc\big|T_{\hc}{}^k_\hga{}^\hbe_j\big|
-K^\hc\big|T_{\hc\ha}{}^\hbe_j\big|
-K_\ha{}^\hc\big|\J_{\hc}{}^\hbe_j
+\J_\ha{}^\hga_jK_\hga{}^\hbe\big|
+\J_\ha{}^\hga_kK^k{}_j\big|~,
\\
\d\phi_\ha{}^{jk}&=&
\Big(\CD_\ha K^\hb\big|-2\ri\J_\ha{}^\hbe_l(\G^\hb)_{\hbe\hga}K^{\hga l}\big|
-K_\ha{}^\hb\big|\Big)\phi_\hb{}^{jk}
+\CD_\ha K^{jk}\big|
+2\phi_\ha{}^{(j}{}_lK^{k)l}\big|
\non\\
&&
+K^\hbe_l\big| R_\ha{}^l_\hbe{}^{jk}\big|
+K^\hb\big| R_{\ha\hb}{}^{jk}\big|
+\J_\ha{}^\hbe_lK^{\hat{C}}\big| R_{\hat{C}}{}_\hbe^l{}^{jk}\big|~,
\\
\d \cV_\ha&=&
\Big(\CD_\ha K^\hb\big|-2\ri\J_\ha{}^\hbe_j(\G^\hb)_{\hbe\hga}K^{\hga j}\big|
-K_\ha{}^\hb\big|\Big)\cV_\hb
+\CD_\ha \tau\big|
+K^\hb\big| F_{\ha\hb}\big|
-2\ri\J_\ha{}^\hbe_jK^j_\hbe\big| ~.~~~~~~~~
\eea
\end{subequations}

\sect{Action  principle in the Wess-Zumino gauge}
\label{WZaction}

Our goal in this section is to reduce the locally supersymmetric action (\ref{InvarAc})
to components in the WZ gauge.  
Using considerations based on eqs. (\ref{InvarAc2}), (\ref{flatac}), (\ref{WZgauge1-1})
and $E|=e$, one can argue that in the WZ gauge it holds
\bea 
 S(\cL^{++}) = S_0 ~+~\dots~, \qquad 
 S_0
&=&
{1\over 2\pi}\oint {(u^+\rd u^{+})\over (u^+u^-)^4}
\int\rd^5 x \,e
\,(\cD^-)^4
\cL^{++}(z,u^+)\Big|~,~~~~~~
\label{projectiveAnsatz0}
\eea
where 
\bea
\cD^-_{\hat \a}:= u^-_i \cD^i_{\hat \a}~, \qquad 
(\cD^-)^4:=-{1\over 96}\ve^{\hal\hbe\hga\hde}\cD_\hal^-\cD_\hbe^-\cD_\hga^-\cD_\hde^- ~,
\eea
and the dots in the expression for $S(\cL^{++})$ in (\ref{projectiveAnsatz0})
denote all the terms with at most three spinor derivatives hitting $\cL^{++}$.

By construction, the action (\ref{InvarAc}) is invariant under 
arbitrary projective transformations (\ref{projectiveGaugeVar}). 
It is remarkable  that the requirement of projective invariance allows one
to uniquely restore the action in the WZ gauge by making use of $S_0$, 
as given in (\ref{projectiveAnsatz0}), as the only input. 
Let us start by presenting the result of explicit calculations announced in \cite{KT-Msugra5D}:
\bea
S(\cL^{++})&=&
{1\over 2\pi} \oint_\g {(u^+\rd u^{+})\over (u^+u^-)^4} 
\int \rd^5 x \,e \Bigg{[}
(\cD^-)^4
+ {\ri\over 4}\J^{\hal\hbe\hga-}\cD^-_\hga\cD^-_{\hal}\cD^-_\hbe
-{25\over 24}\ri \,S^{--}(\cD^-)^2
 \non\\
 &&~
-2(\S^{\ha\hb})_{\hbe}{}^\hga\J_{\ha}{}^{\hbe -}\J_{\hb}{}^{\hde-}
\cD_{{[}\hga}{}^{-}\cD_{\hde{]}}{}^{-}
-{\ri\over 4}\f^{\hal\hbe}{}^{--}\cD^-_{\hal}\cD^-_\hbe
+4(\S^{\ha\hb})^{\hal}{}_{\hga}\f_{{[}\ha}{}^{--}\J_{\hb{]}}{}^{\hga -}\cD^-_\hal
\non\\
&&~
-4\,\J^{\hal\hbe}{}_{}{}_\hbe^{-}S^{--}\cD^{-}_\hal
+2\ri \,\ve^{\ha\hb\hc\hm\hn}(\S_{\hm\hn})_{\hal\hbe}
\J_{\ha}{}^{\hal -}\J_{\hb}{}^{\hbe-}\J_{\hc}{}^{\hga-}\cD_\hga^-
+18S^{--}S^{--}
\non\\
&&~
-6\ri \, \ve^{\ha\hb\hc\hm\hn}(\S_{\hm\hn})_{\hal\hbe}\J_{\ha}{}^{\hal -}
\J_{\hb}{}^{\hbe-}\f_{\hc}{}^{--}
+18\ri \, (\S^{\ha\hb})_{\hal\hbe}\J_{\ha}{}^{\hal-}\J_{\hb}{}^{\hbe -}S^{--}
\Bigg{]}\cL^{++}\Big|~,~~~~~~~~~
\label{Sfin-0}
\eea
where $(\cD^-)^2:=\cD^{\hal -}\cD_{\hal}^-$, $S^{--} := S^{ij} u^-_i u^-_j$, 
$\f_{\hat a}{}^{--}:= \f_{\hat a}{}^{ij} u^-_i u^-_j$ 
and $\J_{\ha}{}^{\hbe-}:= \J_{\ha}{}^{\hbe i} u^-_i$.

The remainder of this section is devoted to the derivation of (\ref{Sfin-0}).
Conceptually, our approach below is quite simple.
We start by computing the variation of $S_0$ (\ref{projectiveAnsatz0}) 
under an infinitesimal projective transformation (\ref{projectiveGaugeVar}),
and then  iteratively add new terms to the action 
in order to cancel out all non-zero contributions 
to the variation,  insuring projective invariance in the end.
Technically, the calculation turns out to be quite long.

In the following, we will use the condensed notation:  
\be
{\rm d} \mu^{++}:= {1\over 2\pi}{(u^+\rd u^{+})\over (u^+u^-)^4}
= - {1\over 2\pi}{(\dt{u}^+u^+) \over (u^+u^-)^4}\,{\rm d} t 
~,
\ee
where we have denoted $\dt{f}:= \rd f(t)/ \rd t$, for a function $f(t)$.
Here $t$ is the time parameter along the closed integration
contour $\g =\{u^+_i(t)\}$ in the isotwistor space 
which occurs  in (\ref{projectiveAnsatz0}).
In the integrand of (\ref{projectiveAnsatz0}),  
the isotwistor $u^-_i$ is chosen to be constant (i.e. time-independent) 
and subject to the condition that $u^+(t)$ and $u^-$ form 
a linearly independent basis at each point of the contour $\g$, that is $(u^+ u^-) \neq 0$. 

Concerning the projective transformations (\ref{projectiveGaugeVar}), 
it is obvious that $S_0$ 
(\ref{projectiveAnsatz0}) is invariant under arbitrary scale 
transformations $u^+_i (t) \to c(t)\,u^+_i(t) $, with $c(t) \neq 0$.
The iterative contributions to $S$ should be chosen to automatically respect 
this invariance. 
It is thus only necessary to analyse projective transformations of $u^-$ of the form
\be
u^-_i~\to~\tilde{u}^-_i\,=\,a(t)\,u^-_i+b(t)\,u^+_i(t)~, \qquad a(t) \neq 0~.
\ee
Since both $u^-$ and $\tilde{u}^-$ should be time independent, 
the coefficients should obey the equations:
\bea
\dt{a}=b\,{(\dt{u}^+u^+)\over (u^+u^-)}~, \qquad \dt{b}=-b\,{(\dt{u}^+u^-)\over (u^+u^-)}~.
\label{ode}
\eea
As is obvious, the functional $S_0$ (\ref{projectiveAnsatz0}) is invariant under 
arbitrary scale transformations $u^-_i \to a(t)\,u^-_i $, with $a \neq 0$.
The other contributions to $S$, which we are going to determine, 
should be chosen to automatically respect 
this invariance. 
Therefore, it only remains  to analyse infinitesimal transformations of the form
$\d u^-_i = b(t) u^+_i$, with $b(t)$ obeying the differential equation  (\ref{ode}).
This transformation induces the following variation of $S_0$: 
\bea
\d S_0&=&
\oint {\rm d} \mu^{++}
\int\rd^5 x\,e
\,\Big{[}\d({\cD}^-)^4\Big{]}\cL^{++}\Big|\non\\
&=&-{\ve^{\hal\hbe\hga\hde}\over 96}
\oint
{\rm d} \mu^{++} \,b
\int\rd^5 x\,e
\,\Big{[}
\,3\,\cD^-_{\hal}\cD^-_{\hbe}\{\cD^+_{\hga},\cD^-_{\hde}\}
+2\,\cD^-_{\hal}\{\cD^+_{\hbe},\cD^-_{\hga}\}\cD^-_{\hde}\non\\
&&~~~~~~~~~~~~~~~~~~~~~~~~~~~~~~~~~~~
+\{\cD^+_{\hal},\cD^-_{\hbe}\}\cD^-_{\hga}\cD^-_{\hde}\,\Big{]}\cL^{++}\Big|~.
\label{VarS0.1}
\eea
${}$First of all, this variation has to be transformed. 

Using the completeness relation
\bea
(u^+u^-)\,\d^i_j=
u^{+i}u^-_j-u^{-i}u^+_j~, 
\eea
the (anti-)commutation relations (\ref{generators}), (\ref{covDev2spinor-})
and (\ref{covDev2spinor-2}) can be seen to imply
\begin{subequations}
\bea
{[}J_{kl},\cD^{\pm}_\hal{]}&=&{1\over (u^+u^-)}\Big{[}\,u^{\pm}_{(k}u^-_{l)}\cD^+_\hal
-u^{\pm}_{(k}u^+_{l)}\cD^-_\hal\,\Big{]}~,
\label{5-9a}
\\
\{\cD^+_\hal,\cD^-_\hbe\}&=&
2\ri(u^+u^-)\cD_{\hal\hbe}+R_{\hal}^+{}_{\hbe}^-{}{}^{kl}J_{kl}
+R_{\hal}^+{}_{\hbe}^-{}{}^{\hga\hde}M_{\hga\hde}
+2\ri(u^+u^-)\ve_{\hal\hbe}Z~,~~~
\label{5-9b}
\\
{[}\cD_{\hal\hbe},\cD^\pm_\hga{]}&=&
{1\over (u^+u^-)}T_{\hal\hbe}{}_{\,}{}_{\hga}^\pm{}_{\,}{}^{\hde-}\cD^+_\hde
-{1\over (u^+u^-)}T_{\hal\hbe}{}_{\,}{}_{\hga}^\pm{}^{\,\hde+}\cD^-_\hde
+R_{\hal\hbe}{}_{\,}{}_{\hga}^\pm{}{}^{\,lp}J_{lp}
+R_{\hal\hbe}{}_{\,}{}_{\hga}^\pm{}_{\,}{}^{\hrh\hta}M_{\hrh\hta}
~.~~~~~~~~~
\label{5-9c}
\eea
\end{subequations}
Here we have introduced  the following definitions:
\bea
R_{\hal}^+{}_{\hbe}^-{}{}^{\hga\hde}&:=&R_{\hal}^i{}_{\hbe}^j{}{}^{\hga\hde}u_i^+u_j^-~,
\quad
R_{\hal}^+{}_{\hbe}^-{}{}^{kl}:=R_{\hal}^i{}_{\hbe}^j{}{}^{kl}u_i^+u_j^-
~,~~~~~~~~\non \\
T_{\hat a}{}_{\,}{}_{\hga}^\pm{}_{\,}{}^{\hde \pm}&:=&
T_{\hat a}{}_{\,}{}_{\hga}^k{}_{\,}{}^{\hde l}u_k^\pm u_l^\pm
~,\quad
R_{\hat a}{}_{\,}{}_{\hga}^\pm{}_{\,}{}^{\hrh\hta}:=
R_{\hat a}{}_{\,}{}_{\hga}^k{}_{\,}{}^{\hrh\hta}u_k^\pm~,\quad
R_{\hat a}{}_{\,}{}_{\hga}^\pm{}{}^{\,lp}:=
R_{\hat a}{}_{\,}{}_{\hga}^k{}{}^{\,lp}u_k^\pm~,
\eea
where the torsion and curvature tensors 
are given explicitly in section 
\ref{SectionSugraGeometry}.
In what follows, we often change the basis in the space of iso-tensors
by the rule: $ A^i  \to A^\pm:=A^i u_i^\pm$.

Let us return to  the variation  (\ref{VarS0.1}). 
We evaluate the anticommutators on the right of  (\ref{VarS0.1}) with the aid of 
(\ref{5-9b}). After that, all vector covariant derivative should 
be moved to the left by making use of (\ref{5-9c}), and all SU(2)-generators 
should be moved to the right using  (\ref{5-9a}). If such transformations produce 
a spinor covariant derivative $\cD^+_{\hat \a}$, it should be pushed to the right 
until it hits $\cL^{++}$, and the latter vanishes due to  the analyticity of the Lagrangian,
$\cD^+_{\hat \a}\cL^{++}=0$. We end up with
\bea
\d S_0&=&
-{\ve^{\hal\hbe\hga\hde}\over 96}\oint {\rm d} \mu^{++}\,b
\int\rd^5 x\,e
\,\Bigg{[}\,
12\ri(u^+u^-)\cD_{\hal\hbe}\cD^-_{\hga}\cD^-_{\hde}
+20(u^+u^-)T_{\hal\hbe}{}_{\,}{}_{\hga}^-{}_{\,}{}^{\hrh-}\cD_{\hrh\hde}
\non\\
&&
+2\ri R_{\hal}^+{}_{\,}{}_{\hbe}^-{}_{\,}{}^{--}\cD_{\hga\hde}
+16\ri T_{\hal\hbe}{}_{\,}{}_{\hga}^-{}_{\,}{}^{\hrh+}\cD^-_{{[}\hrh}\cD^-_{\hde{]}}
-{4\over (u^+u^-)}R_{\hal}^+{}_{\,}{}_{\hbe}^-{}_{\,}{}^{+-}\cD^-_{\hga}\cD^-_\hde
\non\\
&&
-4R_{\hal}^+{}_{\,}{}_{\hbe}^-{}_{\,}{}_{\hga}{}^{\hta}\cD^-_{{[}\hde}\cD^-_{\hta{]}}
+10\ri R_{\hal\hbe}{}_{\,}{}_{\hga}^-{}_{\,}{}^{+-}\cD^-_{\hde}
-10\ri(u^+u^-)R_{\hal\hbe}{}_{\,}{}_{\hga}^-{}_{\,}{}_{\hde}{}^{\hta}\cD^-_{\hta}
\non\\
&&
+6\ri(\cD^-_\hal T_{\hbe\hga}{}_{\,}{}_{\hde}^-{}_{\,}{}^{\hrh+})\cD^-_\hrh
-{2\over (u^+u^-)}(\cD^-_{\hal}R_{\hbe}^+{}_{\,}{}_{\hga}^-{}_{\,}{}^{+-})\cD^-_\hde
+2(\cD^-_{\hal}R_{\hbe}^+{}_{\,}{}_{\hga}^-{}_{\,}{}_{\hde}{}^{\hta})\cD^-_{\hta}
\non\\
&&
+\Bigg(\,6R_{\hal}^+{}_{\,}{}_{\hbe}^-{}_{\,}{}^{kl}\cD^-_{\hga}\cD^-_{\hde}
+8(\cD^-_{\hal}R_{\hbe}^+{}_{\,}{}_{\hga}^-{}_{\,}{}^{kl})\cD^-_{\hde}
-16\ri(u^+u^-)R_{\hal\hbe}{}_{\,}{}_{\hga}^-{}_{\,}{}^{kl}\cD^-_{\hde}
+3(\cD^-_{\hal}\cD^-_{\hbe}R_{\hga}^+{}_{\,}{}_{\hde}^-{}_{\,}{}^{kl})
\non\\
&&~~~~~
-6\ri(u^+u^-)(\cD^-_\hal R_{\hbe\hga}{}_{\,}{}_{\hde}^-{}_{\,}{}^{kl})
-10\ri T_{\hal\hbe}{}_{\,}{}_{\hga}^-{}_{\,}{}^{\hrh-}R_{\hrh}^+{}_{\,}{}_{\hde}^-{}_{\,}{}^{kl}
+2\ri T_{\hal\hbe}{}_{\,}{}_{\hga}^-{}_{\,}{}^{\hrh+}R_{\hrh}^-{}_{\,}{}_{\hde}^-{}_{\,}{}^{kl}
\non\\
&&~~~~~
+{1\over (u^+u^-)}R_{\hal}^+{}_{\,}{}_{\hbe}^-{}_{\,}{}^{--}R_{\hga}^+{}_{\,}{}_{\hde}^-{}_{\,}{}^{kl}
-R_{\hal}^+{}_{\,}{}_{\hbe}^-{}_{\,}{}_{\hga}{}^{\hrh}R_{\hrh}^-{}_{\,}{}_{\hde}^-{}_{\,}{}^{kl}
\Bigg)J_{kl}
\,\Bigg{]}\cL^{++}~.~~~~~~~~~
\label{VarL0.3}
\eea

Let us analyze the contributions to the right-hand side of (\ref{VarL0.3}), 
which are proportional to the SU(2)-generators   $J_{kl}$.
It is important to note that all the coefficients in front of $J_{kl}$ 
are homogeneous functions of degree $1$ in the variables $u^+_i$, 
and  of degree $3$ in $u^-_i$. 
This follows from the fact that such terms come 
from the variation $\d(\cD^-)^4$ which results in replacing   
one of the four isotwistors $(u^-)$'s by 
$(b\,u^+)$. Another piece of useful information 
is the fact that the lagrangian $\cL^{++}$ is a projective superfield of weight 2, 
and hence 
\begin{subequations}
\bea
J_{kl}\cL^{++}&=&-{1\over (u^+u^-)}\Big(u^+_{(k}u^+_{l)}D^{--}-2u^+_{(k}u^-_{l)}\Big)\cL^{++}~,\\
{\rd\over \rd t}\cL^{++}
&=&2{(\dt{u}^+u^-)\over (u^+u^-)}\cL^{++}
-{(\dt{u}^+u^+)\over (u^+u^-)}D^{--}\cL^{++}~,\\
(\dt{u}^+u^+)J_{kl}\cL^{++}&=&
u^+_{(k}u^+_{l)}{\rd\over \rd t}\cL^{++}-2{(\dt{u}^+u^-)\over (u^+u^-)}u^+_{(k}u^+_{l)}\cL^{++}
+2{(\dt{u}^+u^+)\over (u^+u^-)}u^+_{(k}u^-_{l)}\cL^{++}~.~~~
\eea
\end{subequations}
The latter result leads to 
\bea
{(\dt{u}^+u^+)\over (u^+u^-)^4}\,b\,J_{kl}\cL^{++}&=&
{\rd\over \rd t}\Bigg{[}b{u^+_{(k}u^+_{l)}\over (u^+u^-)^4}\cL^{++}\Bigg{]}
+4b{(\dt{u}^+u^+)\over (u^+u^-)^5}u^+_{(k}u^-_{l)}\cL^{++} \non \\
&&+b{(\dt{u}^+u^-)\over (u^+u^-)^5}u^+_{(k}u^+_{l)}\cL^{++}
~.~~~
\eea
This implies that, given an operator $\cO^{(kl)}=\cO^{(lk)}$ 
which is an homogenous function of degree $1$ in $u^+_i$ (as in our case), 
the following equation holds
\bea
\oint {\rm d} \mu^{++}\,b\, \cO^{(kl)}J_{kl}\cL^{++}&=&
\oint  {\rm d} \mu^{++}
\Bigg\{
{4b\,\cO^{+-}\over (u^+u^-)}\cL^{++}
+{b\,u^+_{k}u^+_{l}\over (u^+u^-)}
\Big(D^{--}\cO^{(kl)}\Big)\cL^{++}
\Bigg\}~.~~~~~~~~~
\label{JonL}
\eea
Now, it remains 
to make use of the explicit expressions 
for the  torsion and curvature tensors, see eqs. 
(\ref{covDev2spinor-}--\ref{covDev2spinor-3}),
as well as to notice the relations  
\bea
\cD^-_\hal S^{+-}=-{(u^+u^-)\over 4}\Big(3\cF^{-}_\hal+\cN^{-}_\hal\Big)~,~~~
\cD^{\hal-}F_\hal{}^\hbe={5\over 2}\cF^{\hbe-}~,~~~
\cD^{\hal-}N_\hal{}^\hbe={5\over 2}\cN^{\hbe-}~,~~~
\eea
After some computations, one obtains
\bea
\d S_0&=&
\oint \rd \mu^{++}\,b\int\rd^5 x \,e\,\Bigg{[}\,
 {\ri(u^+u^-)\over 4}\cD^{\hal\hbe}\cD^-_{\hal}\cD^-_\hbe
 +{25\ri\over 12} S^{+-}(\cD^-)^2
 \non\\
 &&~~~~~~~~~~~~~~~~~~~~~~~~
 -5\ri (u^+u^-)\cF^{\hal-}\cD^-_\hal
-22\,S^{--}S^{+-}
\,\Bigg{]}\cL^{++}\Big|~.
\label{VarS0}
\eea

An important remark is in order. 
The original variation $\d S_0$ contained numerous contributions 
proportional to $(u^+u^-)\cN^{\hal-}\cD_\hal^-$. 
All such terms have cancelled out. 
Although at first sight such non-trivial cancellations
may appear miraculous, there is a simple explanation for that.
The point is that such contributions   to the projective variation 
of $S$ are impossible to cancel by means of adding some ``counterterms''
to the action. Complete cancellation is the only option compatible 
with projective invariance.

To cancel out the second and fourth terms in (\ref{VarS0}), we add to $S_0$ 
the following functional:
\bea
S_1&=&\oint {\rm d} \mu^{++}\int\rd^5 x\,e\,\Big{[}-{25\ri\over 24} S^{--}(\cD^-)^2+18\,S^{--}S^{--}
\Big{]}\cL^{++}\Big|~,
\label{S1}
\eea
Evaluating the projective variation of $ S_0+S_1 $ gives
\be
\d \Big(S_0+S_1\Big)=
\oint \rd \mu^{++}\,b\int\rd^5 x \,e\,\Bigg{[}\,
 {\ri(u^+u^-)\over 4}\cD^{\hal\hbe}\cD^-_{\hal}\cD^-_\hbe
 -5\ri (u^+u^-)\cF^{\hal-}\cD^-_\hal
\,\Bigg{]}\cL^{++}\Big|~.
\ee

To simplify the last variation, 
we have to start using the relations that hold in WZ gauge of  section 
\ref{WZgauge}.
In particular,  making use of (\ref{WZgauge1-1})
gives
\bea
\d \Big(S_0+S_1\Big)=
\oint \rd \mu^{++}\,b\int\rd^5 x \,e\,\Bigg{[}\,
 {\ri(u^+u^-)\over 4}\CD^{\hal\hbe}\cD^-_{\hal}\cD^-_\hbe
+{\ri\over 4}\J^{\hal\hbe\hga-}{[}\cD^+_\hga,\cD^-_{\hal}\cD^-_\hbe{]}
~~~~~~~~~~~~~~
\non\\
- {\ri\over 4}\J^{\hal\hbe\hga+}\cD^-_\hga\cD^-_{\hal}\cD^-_\hbe
+{\ri\over 4}\phi^{\hal\hbe}{}^{--}\{\cD^+_{\hal},\cD^-_\hbe\}
+{\ri\over 2}\phi^{\hal\hbe}{}^{+-}\cD^-_{\hal}\cD^-_\hbe
 -5\ri (u^+u^-)\cF^{\hal-}\cD^-_\hal
\,\Bigg{]}\cL^{++}\Big|~.~~~~~~
\label{VarS0-1-2}
\eea
Here, 
the operators proportional to the connection $\phi$ 
can be seen to cancel out by adding the functional
\bea
S_2&=&\oint \rd \mu^{++}\int\rd^5 x \,e\, \Bigg{[}\,
-{\ri\over 4}\phi^{\hal\hbe}{}^{--}\cD^-_{\hal}\cD^-_\hbe
 \,\Bigg{]}\cL^{++}\Big|~.
\eea
Futhermore, in order to cancel out the first term in the second line of (\ref{VarS0-1-2}), 
it is necessary to add one more ``counterterm'':
\bea
S_3&=&
\oint \rd \mu^{++}\int\rd^5 x \,e\,\Bigg{[}\,
 {\ri\over 4}\J^{\hal\hbe\hga-}\cD^-_\hga\cD^-_{\hal}\cD^-_\hbe
 \,\Bigg{]}\cL^{++}\Big|~.
\eea
Now, the projective variation of $S_0+S_1+S_2+S_3$ is 
\bea
&&\d \Big(S_0+S_1+S_2+S_3\Big)=
\oint \rd \mu^{++}\,b\int\rd^5 x \,e\,\Bigg{[}\,
 {\ri(u^+u^-)\over 4}\CD^{\hal\hbe}\cD^-_{\hal}\cD^-_\hbe
\non\\
&&~~~+{\ri\over 2}\J^{\hal\hbe\hga-}{[}\cD^+_\hga,\cD^-_{\hal}\cD^-_\hbe{]}
+ {\ri\over 4}\J^{\hal\hbe\hga-}\cD^-_\hga\{\cD^+_{\hal},\cD^-_\hbe\}
 -5\ri (u^+u^-)\cF^{\hal-}\cD^-_\hal
\,\Bigg{]}\cL^{++}\Big|~.~~~~~~
\label{VarS0-1-2-3-4}
\eea

To simplify the variation obtained, we compute the (anti)commutators
in (\ref{VarS0-1-2-3-4}).
In this way,  we will produce
terms with vector covariant derivatives, $\cD_\ha$,  and also terms with 
the Lorentz and SU(2)-generators.
Then we should systematically move all  the covariant derivatives $\cD_\ha$ to the left, 
with the aid of  the algebra of covariant derivatives,  
and finally make  use of the WZ gauge relation (\ref{WZgauge1-1}). 
Similarly, we should systematically move all the generators to the right using  (\ref{JonL}) and 
$M_{\hal\hbe}\cL^{++}=0$.
As a result, eq. (\ref{VarS0-1-2-3-4}) turns into
\bea
&&\d\Big( S_0+S_1+S_2 +S_3\Big)
=\oint \rd \mu^{++}\,b\,\int\rd^5 x \,e\,\Bigg{[}
~{\ri(u^+u^-)\over 4}(\G^\ha)^{\hga\hde}\CD_\ha\cD_\hga^-\cD_\hde^-
\non\\
&&~
-4(u^+u^-)(\S^{\ha\hb})_{\hbe}{}^{\hga}\CD_{{[}\ha}\J_{\hb{]}}{}^{\hbe-}\cD^-_\hga
+4(u^+u^-)(\S^{\ha\hb})_{\hbe}{}^{\hga}(\CD_{{[}\ha}\J_{\hb{]}}{}^{\hbe-})\cD^-_\hga
-12(\S^{\ha\hb})^{\hal}{}_{\hga}\phi_{{[}\ha}{}^{+-}\J_{\hb{]}}{}^{\hga -}\cD^-_\hal
\non\\
&&~
-4(\S^{\ha\hb})_\hga{}^\hbe\J_\ha{}^{\hga-}\J_\hb{}^{\hde+}\cD^-_{{[}\hde}\cD^-_{\hbe{]}}
+4(\S^{\ha\hb})_\hga{}^\hbe\J_\ha{}^{\hga-}\J_\hb{}^{\hde-}\{\cD^+_\hde,\cD^-_\hbe\}
-2(\S^{\ha\hb})_\hga{}^\hbe\J_\ha{}^{\hga-}\J_\hb{}^{\hde+}\{\cD^-_\hde,\cD^-_\hbe\}
\non\\
&&~
+\J^{\hal\hbe\hga-}\Bigg{(}
{7\ri\over 2(u^+u^-)}R_{\hga}^+{}_{\,}{}_{\hal}^-{}_{\,}{}^{+-}\cD^-_\hbe
+{\ri\over (u^+u^-)} R_{\hal}^+{}_{\,}{}_{\hbe}^-{}_{\,}{}^{+-}\cD^-_\hga
+{\ri\over (u^+u^-)}R_{\hga}^-{}_{\,}{}_{\hal}^-{}_{\,}{}^{++}\cD^-_\hbe
\non\\
&&~~~~~~~~~~~~~~~
+{\ri\over 4(u^+u^-)}R_{\hal}^-{}_{\,}{}_{\hbe}^-{}_{\,}{}^{++}\cD^-_\hga 
+{\ri\over 2}R_{\hga}^+{}_{\,}{}_{\hal}^-{}_{\,}{}_{\hbe}{}^{\hrh}\cD^-_\hrh
+T_{\hga\hbe}{}_{\,}{}_{\hal}^-{}_{\,}{}^{\hde+}\cD^-_\hde
-\hf T_{\hal\hbe}{}_{\,}{}_{\hga}^-{}_{\,}{}^{\hde+}\cD^-_\hde
\non\\
&&~~~~~~~~~~~~~~~
-{2\ri\over (u^+u^-)}(\cD^-_\hal R_{\hga}^+{}_{\,}{}_{\hbe}^-{}_{\,}{}^{+-})
-{\ri\over 2(u^+u^-)}(\cD^-_\hal R_{\hga}^-{}_{\,}{}_{\hbe}^-{}_{\,}{}^{++})
+{\ri\over (u^+u^-)}(\cD^-_\hga R_{\hal}^+{}_{\,}{}_{\hbe}^-{}_{\,}{}^{+-})
\non\\
&&~~~~~~~~~~~~~~~
+{\ri\over 4(u^+u^-)}(\cD^-_\hga R_{\hal}^-{}_{\,}{}_{\hbe}^-{}_{\,}{}^{++})
-4 R_{\hga\hbe}{}_{\,}{}_{\hal}^-{}_{\,}{}^{+-}
+2 R_{\hal\hbe}{}_{\,}{}_{\hga}^-{}_{\,}{}^{+-}
\Bigg{)}
\non\\
&&~
 -5\ri (u^+u^-)\cF^{\hal-}\cD^-_\hal
\,\Bigg{]}\cL^{++}\Big|~.
\label{varS-?1-0}
\eea
In the variation obtained, the first three terms can be simplified 
by using some relations that hold in the WZ gauge.
In particular, the first two terms in (\ref{varS-?1-0})  are of the form
$\CD_\ha U^{\hat a}$, for some $U^{\hat a}$,
and can be simplified by using 
the rule for integration by parts (\ref{intPart}).
${}$Furthermore, the third term in (\ref{varS-?1-0}) can be
transformed to the form:
\bea
(\S^{\ha\hb})^{\hal}{}_{\hga}\big(\CD_{{[}\ha}\J_{\hb{]}}{}^{\hga -}\big)&=&
~{5\ri\over 4}\cF^{\hal -}
-{1\over (u^+u^-)}(\S^{\ha\hb})^{\hal}{}_{\hga}\phi_{{[}\ha}{}^{--}\J_{\hb{]}}{}^{\hga +}
+{1\over (u^+u^-)}(\S^{\ha\hb})^{\hal}{}_{\hga}\phi_{{[}\ha}{}^{+-}\J_{\hb{]}}{}^{\hga -}
\non\\
&&
+\hf(\S^{\ha\hb})^{\hal}{}_{\hga}{\cal T}_{\ha\hb}{}^\hc\J_{\hc}{}^{\hga -}
-{1\over (u^+u^-)}(\S^{\ha\hb})^{\hal}{}_{\hga}\J_{{[}\ha}{}^{\hbe +} 
T_{\hb{]}}{}_{\,}{}_\hbe^-{}_{\,}{}^{\hga -}
\non \\
&&
+{1\over (u^+u^-)}(\S^{\ha\hb})^{\hal}{}_{\hga}\J_{{[}\ha}{}^{\hbe -}
T_{\hb{]}}{}_{\,}{}_\hbe^+{}_{\,}{}^{\hga -}
~,~~~~~~~\,
\eea
as a consequence of the  identity (\ref{WZ12}).
Here the space-time torsion is given by eq.  (\ref{WZ11}) 
which can be equivalently rewritten as follows:
\bea
{\cal T}_{\ha\hb}{}^{\hc}&=&
{4\ri\over (u^+u^-)}(\G^\hc)_{\hga\hde}\J_{{[}\ha}{}^{\hga -}\J_{\hb{]}}{}^{\hde+}
~.~~~~~~~~~
\eea

${}$Further calculations lead to 
\bea
&&\d\Big( S_0+S_1+S_2+S_3\Big)=
\non\\
&&
=\oint \rd \mu^{++}\,b\,\int\rd^5 x \,e\,\Bigg{[}\,
2(\S^{\ha\hb})_{\hbe}{}^\hga\J_{{[}\ha}{}^{\hbe +}\J_{\hb{]}}{}^{\hde-}\cD_{{[}\hga}^{-}\cD_{\hde{]}}^{-}
+2(\S^{\ha\hb})_{\hbe}{}^\hga\J_{{[}\ha}{}^{\hbe -}\J_{\hb{]}}{}^{\hde+}\cD_{{[}\hga}^{-}\cD_{\hde{]}}^{-}
\non\\
&&~
+4(\S^{\ha\hb})_{\hbe}{}^{\hga}\J_{{[}\ha}{}^{\hbe-}\J_{\hb{]}}{}^{\hde-}\{\cD^+_\hde,\cD^-_\hga\}
-2(\S^{\ha\hb})_{\hbe}{}^{\hga}\J_{{[}\ha}{}^{\hbe-}\J_{\hb{]}}{}^{\hde+}\{\cD^-_{\hde},\cD^-_{\hga}\}
\non\\
&&~
+4\J^{\hal\hbe}{}_{\hbe}^{+}S^{--}\cD_{\hal}^{-}
+8\J^{\hal\hbe}{}_{\hbe}^{-}S^{+-}\cD_{\hal}^{-}
-4(\S^{\ha\hb})^{\hal}{}_{\hga}\phi_{{[}\ha}{}^{--}\J_{\hb{]}}{}^{\hga +}\cD^-_\hal
-8(\S^{\ha\hb})^{\hal}{}_{\hga}\phi_{{[}\ha}{}^{+-}\J_{\hb{]}}{}^{\hga -}\cD^-_\hal
\non\\
&&~
+16\ri(\S^{\ha\hb})_{\hal\hbe}
(\G^\hc)_{\hga\hde}\J_{{[}\ha}{}^{\hga-}\J_{\hc{]}}{}^{\hde+}\J_{\hb}{}^{\hal-}\cD^{\hbe-}
-8\ri(\S^{\ha\hb})_{\hal\hbe}
(\G^\hc)_{\hga\hde}\J_{{[}\ha}{}^{\hga -}\J_{\hb{]}}{}^{\hde+}\J_{\hc}{}^{\hal -}\cD^{\hbe-}
\non\\
&&~
+9(u^+u^-)(\G^{\ha})^{\hbe\hga}\J_{\ha}{}^{\hal -}{\X}_{\hbe\hal\hga}{}^-
-18(u^+u^-)(\G^{\ha})_\hal{}^{\hga}\J_{\ha}{}^{\hal -}\cF_{\hga}{}^-
\Bigg{]}\cL^{++}\Big|~.
\label{varS-?2-0}
\eea
Note that the term $ -5\ri (u^+u^-)\cF^{\hal-}\cD^-_\hal$ in (\ref{varS-?1-0}), 
which cannot be consistently produced by the variation of any Lagrangian, 
has cancelled out at this point.

The terms in the fourth line of (\ref{varS-?2-0}) can be seen to 
cancel out by adding to the  action the following functional:
\bea
S_4&=&
\oint \rd \mu^{++}\int\rd^5 x \,e\,\Bigg{[}\,
4(\S^{\ha\hb})^{\hal}{}_{\hga}\phi_{{[}\ha}{}^{--}\J_{\hb{]}}{}^{\hga -}\cD^-_\hal
-4\J^{\hal\hbe}{}_{\hbe}^{-}S^{--}\cD_{\hal}^{-}
 \,\Bigg{]}\cL^{++}\Big|~.
\label{S_4}
\eea
In addition, in order 
to cancel out the two terms quadratic in  spinor derivatives $\cD^-$ 
in the second line of (\ref{varS-?2-0}), 
one has to add to $S$ one more ``counterterm''
\bea
S_5&=&
\oint \rd \mu^{++}\int\rd^5 x \,e\,\Bigg{[}
-2(\S^{\ha\hb})_{\hbe}{}^\hga\J_{\ha}{}^{\hbe -}\J_{\hb}{}^{\hde-}\cD_{{[}\hga}^{-}\cD_{\hde{]}}^{-}
\Bigg{]}\cL^{++}\Big|~.
\eea

At this point, we can simplify $\d\Big( S_0+S_1+S_2+S_3 +S_4+S_5\Big)$ 
by computing the remaining 
anticommutators and then using the same strategy as before, that is:
(i) systematically move all the covariant derivatives $\cD_\ha$ to the left ,  
using the algebra of covariant derivatives,  and then we apply the relation (\ref{WZgauge1-1});
(ii) systematically move  to the right all the generators using  (\ref{JonL}) and 
$M_{\hal\hbe}\cL^{++}=0$.
Such calculations give
\bea
&&\d\Big( S_0+S_1+S_2+S_3 +S_4+S_5\Big)=
\non\\
&&
=\oint \rd \mu^{++}\,b\,\int\rd^5 x \,e\,\Bigg{[}
-24\ri(u^+u^-)(\S^{\ha\hb})_{\hal\hga}\J_{{[}\ha}{}^{(\hal -}\J_{\hb{]}}{}^{\hbe)-}
\Big(F_{\hbe}{}^{\hga}+N_{\hbe}{}^{\hga}+{3\over(u^+u^-)}\d_\hbe^\hga S^{+-}
\Big)
\non\\
&&
~+6\ri(u^+u^-)
\ve^{\ha\hb\hc\hm\hn}(\S_{\hm\hn})_{\hal\hbe}\CD_{\hc}\J_{{[}\ha}{}^{(\hal -}\J_{\hb{]}}{}^{\hbe)-}
+12\ri(u^+u^-)\ve^{\ha\hb\hc\hm\hn}(\S_{\hm\hn})_{\hal\hbe}
\J_{\ha}{}^{\hal -}(\CD_{{[}\hb}\J_{\hc{]}}{}^{\hbe-})
\non\\
&&~
+24\ri \ve^{\ha\hb\hc\hm\hn}(\S_{\hm\hn})_{\hal\hbe}\J_{{[}\ha}{}^{(\hal -}\J_{\hb{]}}{}^{\hbe)-}
\phi_\hc{}^{+-}
-6\ri\ve^{\ha\hb\hc\hm\hn}(\S_{\hm\hn})_{\hal\hbe}\J_{{[}\ha}{}^{(\hal -}\J_{\hb{]}}{}^{\hbe)-}
\J_{\hc}{}^{\hga+}\cD_\hga^-
\non\\
&&~
+16\ri(\S^{\ha\hb})_{\hal\hbe}
(\G^\hc)_{\hga\hde}\J_{{[}\ha}{}^{\hga-}\J_{\hc{]}}{}^{\hde+}\J_{\hb}{}^{\hal-}\cD^{\hbe-}
-8\ri(\S^{\ha\hb})_{\hal\hbe}
(\G^\hc)_{\hga\hde}\J_{{[}\ha}{}^{\hga -}\J_{\hb{]}}{}^{\hde+}\J_{\hc}{}^{\hal -}\cD^{\hbe-}
\non\\
&&~
+9(u^+u^-)(\G^{\ha})^{\hbe\hga}\J_{\ha}{}^{\hal -}{\X}_{\hbe\hal\hga}{}^-
-18(u^+u^-)(\G^{\ha})_\hal{}^{\hga}\J_{\ha}{}^{\hal -}\cF_{\hga}{}^-
\Bigg{]}\cL^{++}\Big|~.
\eea
Here the third line can be simplified by using the integration by parts 
(\ref{intPart}) and the equation
\bea
\CD_{{[}\ha}\J_{\hb{]}}{}^{\hga-}&=&\,{\ri\over 4}\cD^{\hga-}F_{\ha\hb}
+{2\ri\over (u^+u^-)}(\G^\hc)_{\hga\hde}\J_{{[}\ha}{}^{\hga-}\J_{\hb{]}}{}^{\hde+}\J_{\hc}{}^{\hga-}
+{1\over (u^+u^-)}\phi_{{[}\ha}{}^{+-}\J_{\hb{]}}{}^{\hga-}
\non\\
&&
-{1\over (u^+u^-)}\phi_{{[}\ha}{}^{--}\J_{\hb{]}}{}^{\hga+}
-{1\over (u^+u^-)}\J_{[\ha}{}^{\hbe+} T_{\hb{]}}{}_{\,}{}_{\hbe}^-{}_{\,}{}^{\hga-}
+{1\over (u^+u^-)}\J_{[\ha}{}^{\hbe-} T_{\hb{]}}{}_{\,}{}_{\hbe}^+{}_{\,}^{\hga-}~,~~~~~~~\,
\eea
which follows from (\ref{WZ12}).

After some computations, the variation becomes
\bea
&&\d\Big( S_0+S_1+S_2+S_3 +S_4+S_5\Big)=
\non\\
&&=\oint \rd \mu^{++}\,b\,\int\rd^5 x \,e\,\Bigg{[}
-36\ri(\S^{\ha\hb})_{\hal\hbe}\Big(\J_{\ha}{}^{\hal -}\J_{\hb}{}^{\hbe-}S^{+-}
+\J_{\ha}{}^{\hal +}\J_{\hb}{}^{\hbe-}S^{--}\Big)
\non\\
&&~
+12\ri\ve^{\ha\hb\hc\hm\hn}(\S_{\hm\hn})_{\hal\hbe}
\Big(\J_{\ha}{}^{\hal -}\J_{\hb}{}^{\hbe-}\phi_{\hc}{}^{+-}
+\J_{\ha}{}^{\hal -}\J_{\hb}{}^{\hbe+}\phi_{\hc}{}^{--}\Big)
\non\\
&&~
+24\ve^{\ha\hb\hc\hm\hn}(\S_{\hm\hn})_{\hal\hbe}(\G^\hd)_{\hga\hde}
\Big(\J_{\ha}{}^{\hal -}\J_{\hb}{}^{\hbe-}\J_{{[}\hd}{}^{\hga-}\J_{\hc{]}}{}^{\hde+}
+\J_{\ha}{}^{\hal -}\J_{\hb}{}^{\hga-}\J_{\hd}{}^{\hbe-}\J_{\hc}{}^{\hde+}\Big)
\non\\
&&~
-6\ri\ve^{\ha\hb\hc\hm\hn}(\S_{\hm\hn})_{\hal\hbe}
\J_{{[}\ha}{}^{(\hal -}\J_{\hb{]}}{}^{\hbe)-}\J_{\hc}{}^{\hga+}\cD_\hga^-
+16\ri(\S^{\ha\hb})_{\hal\hbe}
(\G^\hc)_{\hga\hde}\J_{{[}\ha}{}^{\hga-}\J_{\hc{]}}{}^{\hde+}\J_{\hb}{}^{\hal-}\cD^{\hbe-}
\non\\
&&~
-8\ri(\S^{\ha\hb})_{\hal\hbe}
(\G^\hc)_{\hga\hde}\J_{{[}\ha}{}^{\hga -}\J_{\hb{]}}{}^{\hde+}\J_{\hc}{}^{\hal -}\cD^{\hbe-}
\Bigg{]}\cL^{++}\Big|~.
\label{varS?3}
\eea
To cancel out the expressions in the second and third lines  of (\ref{varS?3}), 
we have to add to the action the following functional: 
\be
S_6=\oint \rd \mu^{++}\int\rd^5 x \,e\Bigg{[}
\J_{\ha}{}^{\hal-}\J_{\hb}{}^{\hbe -}\Big(18\ri(\S^{\ha\hb})_{\hal\hbe}S^{--}
-6\ri\ve^{\ha\hb\hc\hm\hn}(\S_{\hm\hn})_{\hal\hbe}\phi_{\hc}{}^{--}
\Big)\Bigg{]}\cL^{++}\Big|~.
\label{S_8}
\ee

Now, let us turn our attention to the three gravitini in (\ref{varS?3}). 
For their analysis, we need two auxiliary resuts. 
First, for any tensor $A_{\ha\hb\hc} =-A_{\hb\ha\hc}$, it holds
\bea
(\S^{\ha\hb})_{\hal\hbe}A_{{}\ha\hc{}\hb}&=&
-{3\over 2}(\S^{\ha\hb})_{\hal\hbe}A_{{[}\ha\hb\hc{]}}
+\hf(\S^{\ha\hb})_{\hal\hbe}A_{{}\ha\hb{}\hc}~,~~~
\label{trick-1}
\eea
Given an antisymmetric tensor,  $A^{\hd\he}=-A^{\he\hd}$,
it holds 
\bea
\ve_{\ha\hb\hc\hd\he}(\G^\ha)_{\hal\hbe}(\S^{\hb\hc})_{\hga\hde}A^{\hd\he}&=&
4\ve_{\hal\hbe}A_{\hga\hde}
-4\ve_{\hal\hga}A_{\hbe\hde}
-4\ve_{\hal\hde}A_{\hbe\hga}
+4\ve_{\hbe\hga}A_{\hal\hde}
+4\ve_{\hbe\hde}A_{\hal\hga}
~.~~~~~~
\eea
With the aid of  these identities,
the contributions proportional to three gravitini in (\ref{varS?3}) 
can be seen to be equivalent to
\be
-2\ri \oint \rd \mu^{++}\,b\,\int\rd^5 x \,e\,
\ve^{\ha\hb\hc\hm\hn}(\S_{\hm\hn})_{\hal\hbe}
\Big(
\J_{\ha}{}^{\hal -}\J_{\hb}{}^{\hbe-}\J_{\hc}{}^{\hga+}
+2\J_{\ha}{}^{\hal+}\J_{\hb}{}^{\hbe-}\J_{\hc}{}^{\hga-}
\Big)\cD_\hga^{-}
\cL^{++}\Big|~.
\ee
These terms identically are cancelled out against  the projective variation of the 
functional
\bea
S_7&=& 2\ri \oint \rd \mu^{++}\int\rd^5 x \,e\,
\ve^{\ha\hb\hc\hm\hn}(\S_{\hm\hn})_{\hal\hbe}
\J_{\ha}{}^{\hal -}\J_{\hb}{}^{\hbe-}\J_{\hc}{}^{\hga-}\cD_\hga^-
\,\cL^{++}\Big|~.
\label{S_9}
\eea

${}$Finally, it can be seen that the terms with four gravitini in (\ref{varS?3}) 
cancel each other.
As a result, we obtain
\be
\d\Big( S_0+S_1+S_2+S_3 +S_4+S_5+S_6+S_7\Big)=0~.
\label{varS?4}
\ee
The action (\ref{Sfin-0})  has been proved to be projective invariant.
There is no need to demonstrate its invariance under the local supersymmetry transformations, 
since (\ref{Sfin-0})   is simply the component form of the locally supersymmetric action (\ref{InvarAc})
in the WZ gauge.

Various supergravity-matter systems correspond to different choices for $\cL^{++}$.
Explicit examples of such dynamical systems are given in \cite{KT-Msugra5D}.
\\

\noindent
{\bf Acknowledgements:}\\
Discussions with Ulf Lindstr\"om and Martin Ro\v{c}ek 
are gratefully acknowledged.
SMK is grateful to the organizers of the Workshop in Geometry and  
Supersymmetry, and the Department of Theoretical Physics at Uppsala University, 
for their hospitality.
This work is supported  in part
by the Australian Research Council and by a UWA research grant.

\appendix

\sect{5D Conventions}
\label{sect5DConvenctions}

Our 5D notations and conventions correspond to \cite{KuzLin}. 
The Minkowski metric is given by $\eta_{\hm\hn}={\rm diag}\{-1,1,1,1,1\}$  ($\hm,\hn=0,1,2,3,5$).
The 5D gamma-matrices $\G_{\hat m} = ( \G_m, \G_5 )$, 
with $m=0,1,2,3$,
are defined by 
\be
\{ \G_{\hat m} \, , \,\G_{\hat n} \} 
= - 2 \eta_{\hat m \hat n} \,
\mathbbm{1}
~, \qquad
(\G_{\hat m} )^\dagger = \G_0 \, \G_{\hat m} \, \G_0 
\ee
are chosen in accordance with 
\bea
(\G_m ){}_{\hat \a}{}^{\hat \b}=
\left(
\begin{array}{cc}
0 ~ &  (\s_m)_{\a\bd} \\
(\tilde{\s}_m)^{\ad \b} ~ & 0  
\end{array}
\right)~, \qquad
(\G_5 ){}_{\hat \a}{}^{\hat \b}=
\left(
\begin{array}{cc}
-{\rm i} \,\d_\a{}^\b~ &  0 \\
0 ~ & {\rm i}\, \d^{ \ad}{}_{\bd}
\end{array}
\right)~,
\label{Gamma1}
\eea
such that $\G_0 \G_1 \G_2 \G_3 \G_5 =
\mathbbm{1}$. 
The charge conjugation matrix, $C = (\ve^{\hat \a \hat \b})$, 
and its inverse, $C^{-1} = C^\dag =(\ve_{\hat \a \hat \b})$ 
are defined by 
\bea
C\,\G_{\hat m} \,C^{-1} = (\G_{\hat m}){}^{\rm T}~,
\qquad 
\ve^{\hat \a \hat \b}=
\left(
\begin{array}{cc}
 \ve^{\a \b} &0 \\
0& -\ve_{\ad \bd}    
\end{array}
\right)~, \quad
\ve_{\hat \a \hat \b}=
\left(
\begin{array}{cc}
 \ve_{\a \b} &0 \\
0& -\ve^{\ad \bd}    
\end{array}
\right)~.
\eea
The antisymmetric matrices $\ve^{\hat \a \hat \b}$ and
$\ve_{\hat \a \hat \b}$ are used to raise and lower the four-component 
spinor indices.

A Dirac spinor, $\J=(\J_{\hat \a}) $, and its Dirac conjugate, 
$\bar \J =({\bar \J}^{\hat \a}) = \J^\dag \,\G_0$, look like
\bea
\J_{\hat \a} =   
\left(
\begin{array}{c}
\j_\a \\
{\bar \f}^{\ad}    
\end{array}
\right)~, \qquad 
{\bar \J}^{\hat \a}= (\f^\a \,, \,{\bar \j}_{\ad})~.
\eea
One can now combine ${\bar \J}^{\hat \a}= (\f^\a , {\bar \j}_{\ad})$ and 
$\J^{\hat \a} = \ve^{\hat \a \hat \b} \J_{\hat \b} =(\j^\a , - {\bar \f}_{\ad} )$ 
into a SU(2) doublet, 
\be
\J^{\hat \a}_i = (\J^\a_i,  -{\bar \J}_{\ad i} ) ~, \qquad
(\J^\a_i)^* = {\bar \J}^{\ad i}~, \qquad 
i = \1 , \2 ~,   
\ee 
with $\J^\a_{\1} = \f^\a $ and $\J^\a_{\2} = \j^\a $.
It is understood that the SU(2) indices are raised and lowered 
by $\ve^{ij} $ and  $\ve_{ij} $, $\ve^{\1 \2} =  \ve_{\2 \1} =1$, 
in the standard fashion: $\J^{\hat \a i} = \ve^{ij} \J^{\hat \a}_j$.
The  Dirac  spinor $\J^i = ( \J^i_{\hat \a}  )$
satisfies the pseudo-Majorana condition
${\bar \J}_i{}^{\rm T} = C \,   \J_i$.
This will be concisely represented as
\be
(\J^i_{\hat \a} )^* = \J^{\hat \a}_i~.
\ee

With the definition $\S_{\hat m \hat n} 
=-\S_{\hat n \hat m} = -{1 \over 4}
[\G_{\hat m} , \G_{\hat n} ] $, the matrices 
$\{ \mathbbm{1}, \G_{\hat m} , \S_{\hat m \hat n} \} $
form a basis in the space of  $4 \times 4$ matrices. 
The matrices $\ve_{\hat \a \hat \b}$ and 
$(\G_{\hat m})_{\hat \a \hat \b}$ are antisymmetric, 
$\ve^{\hat \a \hat \b}\, (\G_{\hat m})_{\hat \a \hat \b} =0$, 
while the matrices $(\S_{\hat m \hat n})_{\hat \a \hat \b}$ 
are symmetric.  
Note that any $4\times 4$ matrix ${\rm B}=({\rm B}_\hal{}^\hbe)$ can be 
represented in the form:
\bea
{\rm B}=B\,
\mathbbm{1}
+B^\hm\,\G_\hm
+\hf B^{\hm\hn}\, \S_{\hm\hn}
~,
~~~~~~~~~~~~~~~~
\non \\
B={1\over 4}\tr  \,{\rm B}~,~~~
B^{\hm}=-{1\over 4}\tr\big( \G^\hm\,{\rm B}\big)~,~~~
B^{\hm\hn}=-\tr\big( \S^{\hm\hn}\,{\rm B}\big)~.
\label{Fierz}
\eea

Given a 5-vector $V^{\hat m}$ and an 
antisymmetric tensor $F^{\hat m \hat n} = -F^{\hat n \hat m}$,
we can equivalently represent  them as the 
bi-spinors $V = V^{\hat m} \,  \G_{\hat m}$
and $F = \hf F^{\hat m \hat n}\, \S_{\hat m \hat n} $
with the following symmetry properties
\bea 
V_{\hat \a \hat \b} &=& -V_{\hat \b \hat \a} ~, 
\quad \ve^{\hat \a \hat \b}\, V_{\hat \a \hat \b} =0~, 
 \qquad \quad
F_{\hat \a \hat \b} = F_{\hat \b \hat \a} ~. 
\eea
The two equivalent descriptions 
$ V_{\hat m} \leftrightarrow V_{\hat \a \hat \b}$ and  
$ F_{\hat m \hat n} \leftrightarrow F_{\hat \a \hat \b}$
are explicitly described as follows:
\bea 
V_{\hat \a \hat \b} = V^{\hat m} \, ( \G_{\hat m})_{\hat \a \hat \b}~,
\quad && \quad 
V_{\hat m}  = -{1 \over 4} \,( \G_{\hat m})^{\hat \a \hat \b}\,
V_{\hat \a \hat \b}~, \non \\
F_{\hat \a \hat \b} = \hf F^{\hat m \hat n} 
(\S_{\hat m \hat n})_{\hat \a \hat \b}~, \quad && \quad 
F_{\hat m \hat n}  = (\S_{\hat m \hat n})^{\hat \a \hat \b} \,
F_{\hat \a \hat \b} ~.
\eea
More generally, it holds
\bea
(\G^\hm)_{\hal\hbe}(\G^\hn)_{\hga\hde}F_{\hm\hn}&=&
2\Big(\ve_{\hal\hga}F_{\hbe\hde}
+\ve_{\hbe\hde}F_{\hal\hga}
-\ve_{\hal\hde}F_{\hbe\hga}
-\ve_{\hbe\hga}F_{\hal\hde}\Big)~.
\label{2G1S}
\eea
These results follow from the identities
\bea
 \ve _{\hat \a \hat \b \hat \g \hat \d} 
&=& \ve_{\hat \a \hat \b} \, \ve_{\hat \g \hat \d}
+ \ve_{\hat \a \hat \g} \, \ve_{\hat \d \hat \b}
+\ve_{\hat \a \hat \d} \, \ve_{\hat \b \hat \g}~,  
\non \\
( \G^{\hat m})_{\hat \a \hat \b}\,
( \G_{\hat m})_{\hat \g \hat \d}
&=&
\ve _{\hat \a \hat \b}\, \ve_{ \hat \g \hat \d}
-2 \ve_{\hat \a \hat \g} \, \ve_{\hat \b \hat \d}
+2\ve_{\hat \a \hat \d} \, \ve_{\hat \b \hat \g}
 ~,
\label{someGamma0}
\eea
which imply
\be
 \ve _{\hat \a \hat \b \hat \g \hat \d} 
=\hf \,( \G^{\hat m})_{\hat \a \hat \b}\,
( \G_{\hat m})_{\hat \g \hat \d}
+\hf \, \ve _{\hat \a \hat \b} \, \ve_{ \hat \g \hat \d} ~,
\ee
with 
$ \ve _{\hat \a \hat \b \hat \g \hat \d} $ the completely 
antisymmetric fourth-rank tensor.
Complex conjugation gives 
\be 
(\ve_{\hat \a \hat \b})^* = - \ve^{\hat \a \hat \b}~,
\qquad 
(V_{\hat \a \hat \b})^* = V^{\hat \a \hat \b}~,
\qquad 
(F_{\hat \a \hat \b})^* = F^{\hat \a \hat \b}~,
\ee
provided   $V^{\hat m}$ and  $F^{\hat m \hat n} $ are real.

We often make use of the completely antisymmetric tensor $\ve_{\ha\hb\hc\hd\he}$
that is normalized as
$
\ve_{01235}=-\ve^{01235}=1
$
and possesses the property
\bea
\ve^{\ha\hb\hc\hd\hm}\ve_{\hm\ha'\hb'\hc'\hd'}&=&
-24\d^{{[}\ha}_{{[}\ha'}\d^{\hb}_{\hb'}\d^\hc_{\hc'}\d^{\hd{]}}_{\hd'{]}}
\,=\,-24\d^{{[}\ha}_{\ha'}\d^{\hb}_{\hb'}\d^{\hc}_{\hc'}\d^{\hd{]}}_{\hd'}
\,=\,-24\d^{\ha}_{{[}\ha'}\d^{\hb}_{\hb'}\d^{\hc}_{\hc'}\d^{\hd}_{\hd'{]}}~.
\eea

It is useful to tabulate the products of several gamma-matrices (\ref{Gamma1}).
Making use of (\ref{Fierz}) gives
\begin{subequations}
\bea
\G^\ha\G^\hb
&=&
-\eta^{\ha\hb}
\mathbbm{1}
-2\S^{\ha\hb}
~,
\label{someGamma1}
\\
\G^\ha\G^\hb\G^\hc
&=&
(-\eta^{\ha\hb}\eta^{\hc\hd}+\eta^{\hc\ha}\eta^{\hb\hd}-\eta^{\hb\hc}\eta^{\ha\hd})\,\G_\hd
+\ve^{\ha\hb\hc\hd\he}\,\S_{\hd\he}
~,~~~
\label{someGamma15}
\\
\G^{\ha}\G^{\hb}\G^{\hc}\G^{\hd}
&=&
(\eta^{\ha\hb}\eta^{\hc\hd}-\eta^{\ha\hc}\eta^{\hb\hd}+\eta^{\ha\hd}\eta^{\hb\hc})\mathbbm{1}
-\ve^{\ha\hb\hc\hd\he}\, \G_\he
+2\eta^{\ha\hb}\, \S^{\hc\hd}
\non\\
&&
-2\eta^{\ha\hc}\,\S^{\hb\hd}
+2\eta^{\hb\hc} \, \S^{\ha\hd}
+2\eta^{\hd\hc} \, \S^{\ha\hb}
-2\eta^{\hd\hb}\, \S^{\ha\hc}
+2\eta^{\hd\ha} \S^{\hb\hc}
~,
\label{someGamma16}
\\
\G^{\ha}\G^{\hb}\G^{\hc}\G^{\hd}\G^\he
&=&
\,\ve^{\ha\hb\hc\hd\he}\mathbbm{1}
+
\, \G^\ha(\eta^{\hb\hc}\eta^{\hd\he}-\eta^{\hb\hd}\eta^{\hc\he}+\eta^{\hc\hd}\eta^{\hb\he})
\non\\
&&
+\G^\hb(-\eta^{\hc\hd}\eta^{\he\ha}+\eta^{\hc\he}\eta^{\hd\ha}-\eta^{\hd\he}\eta^{\hc\ha})
+\G^\hc(\eta^{\hd\he}\eta^{\ha\hb}-\eta^{\hd\ha}\eta^{\he\hb}+\eta^{\he\ha}\eta^{\hd\hb})
\non\\
&&
+\G^\hd(-\eta^{\he\ha}\eta^{\hb\hc}+\eta^{\he\hb}\eta^{\ha\hc}-\eta^{\ha\hb}\eta^{\he\hc})
+\G^\he(\eta^{\ha\hb}\eta^{\hc\hd}-\eta^{\hc\ha}\eta^{\hb\hd}+\eta^{\hb\hc}\eta^{\ha\hd})
\non\\
&&
+2\ve^{\ha\hb\hc\hd\hm} \, \S_\hm{}^\he
-\eta^{\ha\hb}\ve^{\hc\hd\he\hm\hn}\,\S_{\hm\hn}
+\eta^{\hc\ha}\ve^{\hb\hd\he\hm\hn} \, \S_{\hm\hn}
-\eta^{\hb\hc}\ve^{\ha\hd\he\hm\hn}\, \S_{\hm\hn}
\non\\
&&
-\eta^{\hd\ha}\ve^{\hb\hc\he\hm\hn} \, \S_{\hm\hn}
+\eta^{\hd\hb}\ve^{\ha\hc\he\hm\hn} \, \S_{\hm\hn}
-\eta^{\hd\hc}\ve^{\ha\hb\he\hm\hn} \, \S_{\hm\hn}
~.
\label{someGamma19}
\eea
\end{subequations}

In conclusion, we give a useful relation often used in the paper. It is
\bea
\ve_{\ha\hb\hc\hd\he}(\G^\hc)_{\hal\hbe}(\S^{\hd\he})_{\hga\hde}
&=&
2\ve_{\hal\hbe}(\S_{\ha\hb})_{\hga\hde}
+2\ve_{\hga\hal}(\S_{\ha\hb})_{\hbe\hde}
+2\ve_{\hde\hal}(\S_{\ha\hb})_{\hbe\hga}
\non\\
&&
-2\ve_{\hga\hbe}(\S_{\ha\hb})_{\hal\hde}
-2\ve_{\hde\hbe}(\S_{\ha\hb})_{\hal\hga}~.
\eea

\small{

}

\end{document}